\documentclass[a4paper,11pt]{article}
\usepackage[utf8]{inputenc}
\usepackage[english]{babel}

\usepackage{graphicx}
\graphicspath{ {./images/} }

\usepackage[margin=1in]{geometry}
\usepackage{authblk}
\usepackage{csquotes}
\usepackage{biblatex}
\usepackage{amsmath}
\usepackage{mathtools}
\usepackage{algorithm}
\usepackage{bm}
\usepackage{float}
\usepackage{pdfpages}

\usepackage[hidelinks]{hyperref}
\hypersetup{
    colorlinks = true,
    linkcolor  = blue
}

\renewenvironment{abstract}
 {\small
  \begin{center}
  \bfseries \abstractname\vspace{-.5em}\vspace{0pt}
  \end{center}
  \list{}{
    \setlength{\leftmargin}{.5cm}%
    \setlength{\rightmargin}{\leftmargin}%
  }%
  \item\relax}
 {\endlist}

\DeclareMathOperator*{\argmin}{arg\,min}

\begin{document}

\begin{titlepage}
    \begin{center}
        \vspace*{1cm}
            
        \Huge
        \textbf{Stochastic Epidemic Modelling}
            
        \vspace{0.5cm}
        \LARGE
        MATH32200 - 20cp Project
            
        \vspace{1.5cm}
            
        \textbf{Georgios Efstathiadis}\\
        \vspace{0.5cm}
        \LARGE
        Supervised by Dr. Anthony Lee
        \\ 
            
        \vfill
            
        A thesis presented for the degree of\\
        BSc Mathematics with Statistics
            
        \vspace{0.8cm}
            
        \includegraphics[width=0.4\textwidth]{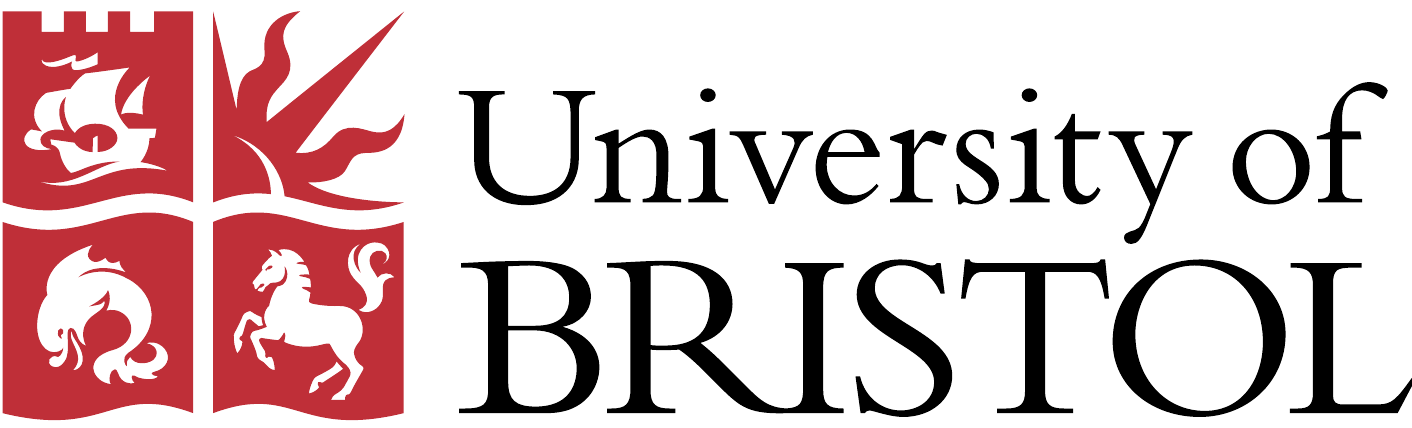}
            
        \Large
        Department of Mathematics\\
        University of Bristol\\
        United Kingdom\\
        May 9, 2022
            
    \end{center}
\end{titlepage}
\pagenumbering{roman}

\includepdf[pages=-]{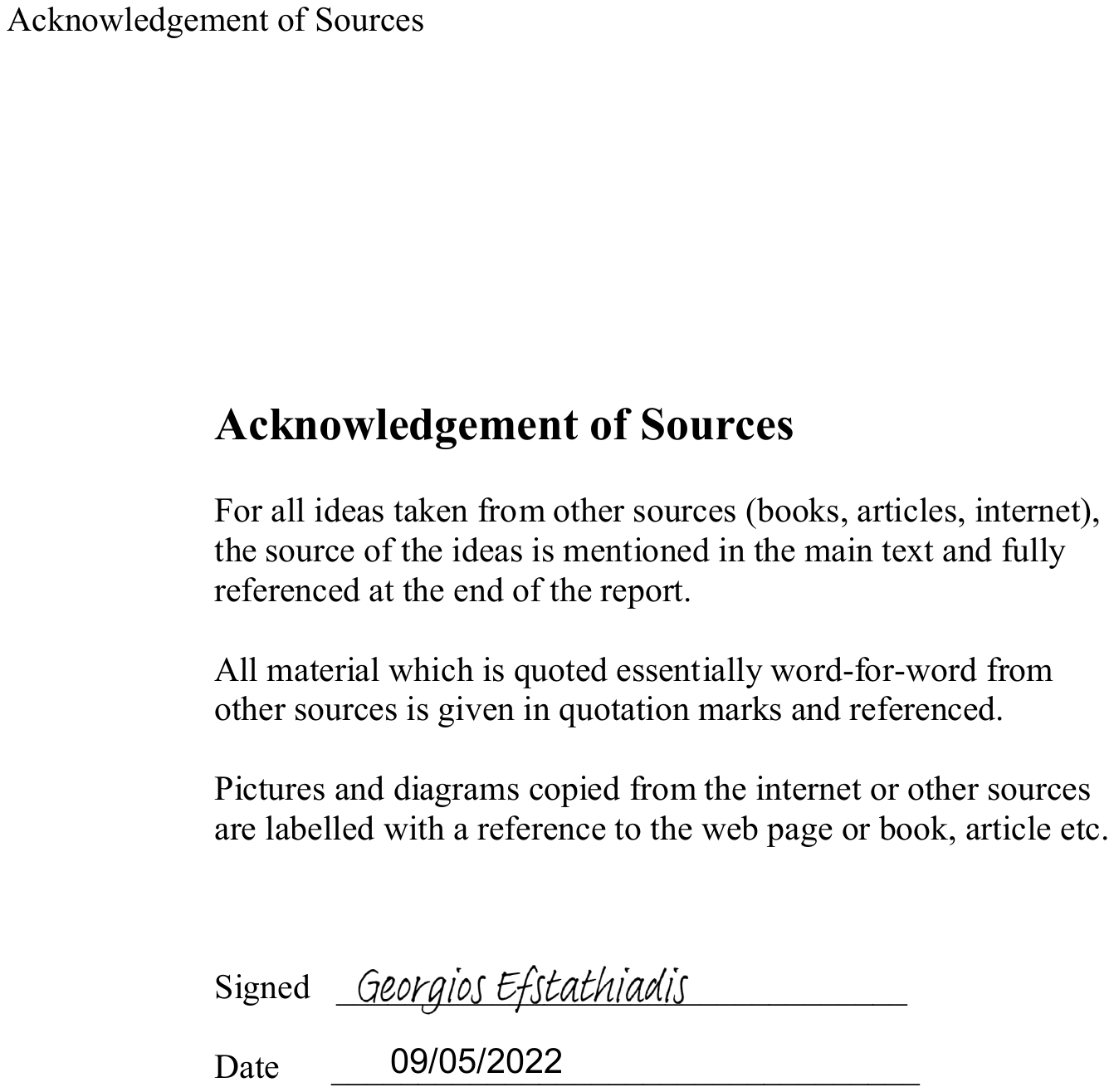}

\tableofcontents
\thispagestyle{empty}
\newpage

\pagenumbering{arabic}

\begin{abstract}

    Inferring how an epidemic will progress and what actions to take when presented with limited information is of critical importance for epidemiologists and health professionals. In real world settings, epidemiology data can be scarce or subject to reporting errors. In this project there are different epidemic scenarios simulated and, using hidden Markov Chains, it is attempted to mimic the imperfect data an epidemiologist will encounter. Furthermore, different kinds of compartmental models are modelled using the particle Markov Chain Monte Carlo algorithm with a variation of the adaptive Metropolis-Hastings algorithm to estimate the posterior density of the parameters underlying the models. Moreover, the sensitivity of these algorithms is investigated when subjected with changes in the dataset. This is accomplished by limiting the information provided, while using an adaptive approach on the posterior covariance of the parameters.

\end{abstract}

\section{Introduction}

In an epidemic setting it is common to be interested in the parameters underlying the disease, in order to be able to make health policy decisions on how to contain the epidemic or create a timeline of how it is expected to evolve. Compartmental models are a relatively simple and quick way to get some estimates for the disease's characteristics on a population level and make informed decisions on how to react, which were also used in the beginning of the COVID-19 pandemic  (\cite{adam-2020}, \cite{weissman-2020}). In this project the use of the Particle Markov Chain Monte Carlo (PMCMC) algorithm is explored, in modelling these stochastic epidemic models. The use of PMCMC in epidemic modelling has been investigated before, but not in the same manner as here. Dureau and others \cite{dureau-2013} researched the use of PMCMC in the SEIR model, with some variations, and time-varying parameters. Rasmussen and others \cite{rasmussen-2011} created a framework relying on PMCMC to combine compartmental models and genetics data to obtain the underlying disease dynamics and more recently Endo and others \cite{endo-2019} presented a paper that acts as a tutorial for anyone who wants to apply the PMCMC algorithm to model epidemics with compartmental models.

In Section 3, the methodology is presented -the theory behind the compartmental models, specifically the SIR model, as well as the algorithms used to simulate stochastic epidemic data and estimate their parameter's posterior densities. In Section 4, the results of fitting different scenarios using Approximate Bayesian Computation (ABC) and PMCMC are presented, namely results from modelling SIR and SEIR compartmental models with hidden Markov Chains to simulate real world scenarios. Two real world scenarios are looked into, noisy data and under-reported data which are simulated with the Normal and the Binomial distributions respectively. After that, some experiments are reported on whether it is still possible to obtain Markov Chains that provide us with good posterior samples, while limiting the information in the dataset and reducing the amount of data the model trains on. It is attempted to limit the simulated data by increasing the noise in the noisy data and decreasing the probability of observation of cases in the under-reported data. Moreover, also investigated is what happens when trying to infer the parameters underlying those models earlier in the disease's progression, as it is imperative for healthcare professionals to be able to understand the disease's characteristics as early as possible. Finally, in the last section, Section 5, the possible extensions of this work are discussed as well as ideas worked on in this project that did not provide any meaningful results. Additionally, mentioned are some of the limitations of this algorithm in an epidemiology setting and some of the difficulties encountered.

\section{Methodology} \label{sec:meth}

\subsection{Epidemiological Compartmental Models} \label{sec:compartment}

When dealing with epidemics that occur within a limited time-frame, so that outside forces affecting the population (births and deaths) are negligible, a common and simple model to use is the SIR without demographics by Kermack WO and McKendrick AG \cite{kermack-1991}. This simple model can be split in 3 compartments: \textit{S} which stands for Susceptible, the population that can contract the disease, \textit{I} which stands for Infected, the population that can infect others, and \textit{R} which stands for Removed, the population that has either recovered or died of the disease and ,thus, can no longer contract it. In this simple setting we assume that all contacts between any compartment are homogeneous and that people who belong in the same compartments adhere to the same disease dynamics. Hence we say that at time \textit{t}, \textit{S(t), I(t), R(t)} define the number of people who belong in each compartment. Since our model is a closed system we have that 

\[ S(t) + I(t) + R(t) = N \ \textrm{for all}\ t \]

where $N$ is the total population, which is constant across time.

The system is defined by a set of three equations:
\begin{equation}
\begin{split}
&S'(t) = -\beta S(t) I(t) \\
&I'(t) = \beta S(t) I(t) - \gamma I(t) \\
&R'(t) = \gamma I(t)
\end{split}
\end{equation}

Where $\beta$ represents the rate at which people leave the Susceptible class to enter the Infected class and $\gamma$ the rate at which people leave the Infected class to enter the Removed class. A way to interpret $\beta$ and $\gamma$ in the epidemics setting is that $\beta$ is the rate of contact of a susceptible person times the probability of infection and $1/\gamma$ is the average time it takes an infected person to recover or die from the disease.

The last line in equation (1) is redundant, since we can simply find $R(t)$ at each time-step from the other two equations using $R(t) = N - I(t) - S(t)$. Hence the system can be described by:

\begin{equation}
\begin{split}
&S'(t) = -\beta S(t) I(t) \\
&I'(t) = \beta S(t) I(t) - \gamma I(t)
\end{split}
\end{equation}

We say that the disease has run its course when at $t^\star$ we have $S(t^\star) = 0$, meaning that everybody susceptible to the epidemic has been actually infected, or when $I(t^\star) = 0$, meaning there are no more infected individuals. By looking at the equations above we can see that $I'(t) > 0$ if and only if $\beta S(t) - \gamma > 0$. Since we need $I'(0) > 0$ in order to have an epidemic we need $S(0) > \gamma / \beta$. We define $R_0\coloneqq \beta N / \gamma$, when $I(0)$ is very small compared to $S(0)$, as the \textit{basic reproduction number} which is critical to whether the disease is going to progress to an epidemic or it is going to die out. If $R_0<1$ the disease will finish progressing with people still not infected, while if $R_0>1$ the disease will finish progressing when everybody has been infected.

There are many different compartmental models used depending on the nature of the disease. Another model I will investigate is the SEIR model. In the SEIR there is an extra compartment, $E$, namely the Exposed population. If a person is in the exposed compartment they have contracted the disease, but they are not yet infectious, meaning they cannot infect others. Since our model remains a closed system, births and deaths of natural causes are insignificant in the time it takes the epidemic to progress, we have $S(t) + E(t) + I(t) + R(t) = N\ \textrm{for all}\  t$, The SEIR model is defined by the set of the following equations:

\begin{equation}
\begin{split}
&S'(t) = -\beta S(t) I(t) \\
&E'(t) = \beta S(t) I(t) - \alpha E(t) \\
&I'(t) = \alpha E(t) - \gamma I(t) \\
&R'(t) = \gamma I(t)
\end{split}
\end{equation}

In the SEIR model we have an extra parameter, $\alpha$, which is the rate of leaving the exposed compartment to enter the infectious one. It can be interpreted as $1/\alpha$ signifying the average period a person spends in the exposed class, or the \textit{latent period}. One thing to be noted about the SEIR model is that to enter the infectious compartment, you first need to enter the exposed one, so you cannot be infectious immediately after you contact the disease.

Having defined the deterministic SIR model we can now define the stochastic SIR model which progresses in continuous time \cite{greenwood-2009}. The system can now be described by a Markov jump process in continuous time where there are two types of jumps, $(S, I) \rightarrow (S-1, I+1)$, when a person gets infected, and $(S, I) \rightarrow (S, I-1)$, when a person recovers or dies. These events occur at a categorical probability with the times in between each jump being identically and independently exponentially distributed. Thus, in a time interval $[t, t + \Delta t]$ we have the following probabilities of a jump happening that define the stochastic SIR model:

\begin{equation}
\begin{split}
&P((S_{t+\Delta t},I_{t+\Delta t}) - (S_{t},I_{t}) = (-1,1)) = \beta \frac{S_t I_t}{N}\Delta t + o(\Delta t) \\
&P((S_{t+\Delta t},I_{t+\Delta t}) - (S_{t},I_{t}) = (0,-1)) = \gamma I_t \Delta t + o(\Delta t) \\
&P((S_{t+\Delta t},I_{t+\Delta t}) - (S_{t},I_{t}) = (0,0)) = 1 - (\beta \frac{S_t}{N} + \gamma) I_t \Delta t + o(\Delta t)
\end{split}
\end{equation}

The process $(S_t, I_t)$ for $t \geq 0$ is a Poisson process where the \textit{stochastic rate} $\lambda (t)$ depends on time and the state of the process at time $t$. The jump $(S, I) \rightarrow (S-1, I+1)$ occurs at a stochastic rate $\beta \frac{S_t I_t}{N}$ and the jump $(S, I) \rightarrow (S, I-1)$ occurs at a stochastic rate $\gamma I_t$. We can clearly see that the probabilities for a jump to occur at a time interval $[t, t + \Delta t]$ are conditional at the state of the system at time $t$.

\subsection{Gillespie Algorithm}

To simulate a stochastic SIR model I will be using the Gillespie Algorithm. The algorithm, created by Jacob L. Doob \cite{doob-1945} and popularised by Daniel T. Gillespie in 1976 \cite{gillespie-1976}, is used to simulate stochastic trajectories of a system where the rates of the random variables are known. Before we define the algorithm, we need to define the propensity functions of a stochastic system. These are the stochastic rates of the jumps in the system that were defined previously.

The direct formulation of the algorithm is as follows \cite{gillespie-2007}:
\begin{algorithm}
\caption{Gillespie Algorithm}
\begin{enumerate}
  \item System is initialised and we have $t=t_0$, starting time, and $\bm{x}=\bm{x_0}$, starting population, known.
  \item Using $a_j$ as the propensity functions of the system evaluate $a_j(\bm{x})$ for each possible action $j$.
  \item Draw $r_1$, $r_2$ uniform random variables on $[0, 1]$.
  \item Calculate $\tau = \frac{1}{\sum a_j(\bm{x})} log(\frac{1}{r_1})$, sampled exponentially distributed time, and $j = \argmin_x (\sum_{j'=1}^{h}a_{j'}(\bm{x}) > r_2 \sum a_{j'}(\bm{x}))$, sampled from categorical distribution over possible events.
  \item Replace $t \leftarrow t + \tau$ and $\bm{x} \leftarrow \bm{x} + \bm{v}_j$, where $\bm{v}_j$ is the change of population when action $j$ occurs.
  \item Return to step 2 or end simulation.
\end{enumerate}
\end{algorithm}

In the case of the stochastic SIR model, where $x = (S, I, R)$, we just replace $a_j$ with:
\begin{equation}
\begin{split}
&a_0(\bm{x}) = \beta \frac{S I}{N} \\
&a_1(\bm{x}) = \gamma I
\end{split}
\end{equation}

and $v_j$ with:
\begin{equation}
\begin{split}
&v_0 = (-1, +1, 0) \\
&v_1 = (0, -1, +1)
\end{split}
\end{equation}

\subsection{Markov Chain Monte Carlo}

When trying to model an infectious disease with a stochastic SIR model the parameters of the model are going to be unknown. One of the ways  we can estimate those parameters is Markov Chain Monte Carlo or MCMC - a class of computational methods - which in Bayesian Statistics are used to sample from the posterior distribution of some random variable. The term MCMC can be split into two parts, \textit{Monte Carlo} and \textit{Markov Chain}. Monte Carlo is another class of algorithms which are used to sample large amounts of random variables and get estimates of statistical properties for these random variables that resemble the real ones. Markov Chain is a stochastic process defined by a sequence or "chain" of events that happen in a set of time, which can be discrete or continuous. Markov Chains are characterised by the fact that they are "memory-less", meaning that the probability of being at a state, $X_{n+1}$, only depends on the previous state, $X_n$,  and not on the ones that came before it. Hence a discrete Markov Chain is defined by the following probability equation, also known as Markov ”memory-less” property:

\begin{equation}
P(X_{n+1} = x | X_1 = x_1, X_2 = x_2, ..., X_n = x_n) = P(X_{n+1} = x | X_n = x_n)
\end{equation}

Having defined the parts that comprise MCMC algorithms we can now describe how a typical MCMC algorithm works. MCMC algorithms are used to conduct bayesian inference so, given some realisation of a random variable $X$, they try to sample from the posterior distribution of a random variable $Y$. The Bayes theorem states that:

\begin{equation}
P(Y = y|X = x) = \frac{P(X = x|Y = y) P(Y = y)}{P(X = x)}
\end{equation}

where $P(Y = y|X = x)$ is the posterior distribution of $Y$, $P(X = x|Y = y)$ is the likelihood that data $X$ are true for the given $Y$ and $P(Y = y)$ is the prior distribution of $Y$. If we define function $p(y; x) = L(y; x) \pi(y)$, where $L(y; x) = P(X = x|Y = y)$ and $\pi(y) = P(Y = y)$, then we have

\begin{equation}
\label{eq:bayes_prop}
P(Y=y|X=x) \propto p(y; x)
\end{equation}

up to a multiplication constant that does not depend on $y$, also known as the \textit{normalising constant}.

It is very common that in many applications we may not know how to calculate the marginal probability distribution function of $X$, thus we cannot calculate the posterior distribution analytically. MCMC algorithms are used to construct intervals that are then used to estimate the posterior density via bayes rule and they have the advantage that you only need to know the target probability distribution function up to a normalising constant. MCMC algorithms work by generating samples $\theta_{t+1}$ from a \textit{proposal distribution} $h(\cdot | \theta_t)$ that depends only on the last sample generated (Markov "memory-less" property). Thus, they generate chains that after enough samples have been generated converge to the posterior distribution of $\theta$ (Monte Carlo simulation). In the specific MCMC algorithm we are going to be using, Metropolis-Hastings algorithm, a generated sample is accepted to the chain under a given probability, the \textit{acceptance probability}, that depends on the proposal distribution $h(\cdot)$ and the target distribution $P(\cdot)$ which is proportional to the likelihood function times the prior for some given data $X$. If the generated sample of $\theta$ is not accepted, then the previously accepted sample is added to the posterior distribution instead.

Metropolis-Hastings algorithm \cite{hastings-1970} is one of the most common MCMC algorithms and is a generalisation of the Metropolis algorithm which is actually the first MCMC algorithm \cite{metropolis-1953}.

\begin{algorithm}
\caption{Metropolis-Hastings Algorithm}
\begin{enumerate}
  \item System is initialised and we have $i=0$, starting time, and sample from prior distribution $\theta_0 \sim v$.
  \item Using the proposal distribution we sample $\theta' \sim h(\cdot | \theta_i)$ and calculate the probability of acceptance $p_i = \min(1, \frac{P(\theta')h(\theta_i|\theta')}{P(\theta_i)h(\theta'|\theta_i)})$.
  \item With probability $p_i$ we accept $\theta'$ and $\theta_{i+1} \leftarrow \theta'$ and with probability $1-p_i$ we discard $\theta'$ and $\theta_{i+1} \leftarrow \theta_i$.
  \item Increment $i$ by 1 and return to step 2, or finish loop if a set number of samples from $\theta$ has been generated.
\end{enumerate}
\end{algorithm}

We can rewrite the probability of acceptance from the Metropolis-Hasting algorithm using the fact that the target distribution is proportional to the likelihood times the prior.

\begin{equation}
\begin{split}
p_i
&= \min(1, \frac{P(\theta')h(\theta_i|\theta')}{P(\theta)h(\theta'|\theta_i)}) \\
&= \min(1, \frac{L(\theta'; X)v(\theta')h(\theta_i|\theta')}{L(\theta_i; X)v(\theta)h(\theta'|\theta_i)})
\end{split}
\end{equation}

Thus, by Equation \ref{eq:bayes_prop} we get that a simulated $\theta'$ is more likely to be accepted the higher the posterior probability of $\theta'$ given some data $X$ compared to the posterior probability of the previously sampled $\theta_i$. Additionally, if the proposal distribution is symmetric, then $h(\theta_i|\theta')=h(\theta_i|\theta')$ and the two terms cancel out. If the prior distribution is proportional to a constant, we also have $v(\theta')=v(\theta_i)$ and these two terms cancel out as well. Then the probability of acceptance depends only on the likelihood.

\begin{equation}
p_i = \min(1, \frac{L(\theta'; X)}{L(\theta_i; X)})
\end{equation}

Though MCMC is a very useful set of algorithms it is hard to determine the proposal distribution and the prior distribution (initial conditions) to ensure convergence is achieved in a relatively small amount of steps and it is thus a computationally intensive process.

\subsection{Approximate Bayesian Computation}

Another way to sample from posterior distributions of parameters is Approximate Bayesian Computation or ABC \cite{rubin-1984}. ABC is a class of Bayesian computational methods used in parametric inference which, given a prior distribution of a set of parameters and some data, can estimate the posterior distribution with the advantage, compared to MCMC, that you do not need to know the posterior density up to a multiplication constant.

There are many algorithms included in the ABC class of algorithms, but the simplest one is the rejection sampling algorithm \cite{csillery-2010}. In the rejection sampling algorithm given some model $g(\cdot | \theta)$ that simulates data depending on a parameter $\theta$ and a prior distribution for $\theta$ we draw $n$ samples $\theta$ from the prior distribution and then use them to simulate data $X_i$ from $g(\cdot | \theta_i)\  \forall\ i \in [1, n]$. If some selected summary statistics of $X_i$ are close enough to the summary statistics of the real data $D$, given a distance function, we accept $\theta_i$ in the posterior distribution, otherwise we discard it. Thus, the algorithm requires summary statistics to compare on, a distance function to calculate how similar are the simulated and real data, as well as a threshold $\epsilon$ such that if distance between real and simulated data is less than threshold $\epsilon$ the sampled theta is accepted.

\begin{algorithm}
\caption{Rejection Sampling Algorithm}
\begin{enumerate}
  \item Sample $\theta_i \sim v$, starting with $i=1$.
  \item Sample $X_i  \sim g(\cdot | \theta_i)$ and then calculate $S(X_i)$ for some summary statistics function $S$.
  \item Calculate the distance between the summary statistics of the simulated and real data $h(S(X_i), S(D))$, where $h$ is the distance function and $D$ the observed data.
  \item If the distance is less or equal than threshold $\epsilon$, accept $\theta_i$ otherwise do nothing.
  \item Increment $i$ by $1$ and return to step 2, or end loop if a set number of samples have been accepted and the posterior distribution has been recorded.
\end{enumerate}
\end{algorithm}

 The estimated posterior distribution of $\theta$ from the rejection sampling algorithm will look like the following:

\begin{equation}
\begin{split}
&\theta \sim \frac{v(\cdot) \int g(x | \cdot) I(h(S(x, S(D)) \leq \epsilon) dx}{Z} \\
&Z = \int v(\theta) \int g(X_{\theta} | \theta) I(h(S(X_{\theta}, S(D)) \leq \epsilon) dX_{\theta} d\theta
\end{split}
\end{equation}

where $I$ is the indicator function. The exact posterior distribution of $\theta$ is $\theta \sim v(\cdot) f(S(D) | \cdot)$. Thus, it is clear that when $\epsilon \rightarrow 0$ the approximated posterior distribution we get from ABC converges to the exact one.

In the case of the stochastic SIR model we can use a rejection sampling algorithm, by replacing $g$ above with the Gillespie algorithm for SIR models, $S$ with something like the number of infected and susceptible populations at each discrete time-point and $h$ with the absolute difference of its arguments.

\subsection{Sequential Monte Carlo}

Sequential Monte Carlo are a set of Monte Carlo algorithms used to sample the posterior states of a Markov process \cite{doucet-2006}. Here we are going to use Sequential Monte Carlo, or Particle Filters as they are called, in the context of discrete time hidden Markov Chains. A hidden Markov Chain is described by a set of states $X_{0:n} \coloneqq (X_0, ..., X_n)$, which are themselves a Markov Chain, but are not observed, and a set of states $Y_{0:n-1} \coloneqq (Y_0, ..., Y_{n-1})$, which are observed. Hidden Markov Models depend on the following distributions.

\begin{equation}
\label{eq:hidden1}
\begin{split}
&X_0 \sim \mu\ \textrm{and}\ X_p \sim g(\cdot | X_{p-1})\ \textrm{for}\ p\ \textrm{in}\ [1, n] \\
&Y_p \sim g^y(\cdot | X_p)\ \textrm{for}\ p\ \textrm{in}\ [0, n-1]
\end{split}
\end{equation}

where $\mu$ is the initial distribution of $X_0$, $X_p$ is sampled from a distribution that depends only on the previous state $X_{p-1}$ and $Y_p$ is sampled from another distribution that depends only on the Markov Chain at time $p$ which is $X_p$.
In the context of Hidden Markov Models some interesting statistics we want to retrieve are the likelihood of the observations $Y_p$ for $p \in \{0,n-1\}$ and get samples of the hidden Markov Chain $X_p$ for $p \in \{0, n\}$.

Concerning the likelihood of the observations there are ways to get the exact values analytically, but only for a few special cases. A more feasible way to estimate the likelihoods is using a \textit{Particle Filter} (\cite{doucet-2000}, \cite{doucet-2006}). A particle filter works by generating particles at each time-step from 0 to $n$ that simulate $X_p$, the hidden chain, and then calculate their weights according to how likely the particles are to match the actual $X_p$ given the observations $Y_p$. Then in the next time-step only the particles with the highest weights are selected to sample from the next set of particles. The thinking behind particle filters is that at each time-step the system corrects itself by only choosing the best particles according to the observations. This can be less computationally intensive than other methods, since it requires less simulations to achieve the same or better results.

\begin{algorithm}
\caption{Particle Filter}
\begin{enumerate}
\label{algo:particle_filter}
  \item Initiate the algorithm by setting $Z_0 = 1$, setting time $t=1$ and sampling $N$ particles $\zeta_0^i \sim \mu$ where $\mu$ is the initial distribution of $X_0$.
  \item Set $Z_t \leftarrow Z_{t-1}\frac{\sum_{i=1}^N g^y(Y_{t-1} = y_{t-1} | X_{t-1} = \zeta_{t-1}^i)}{N}$.
  \item Let $w_i=\frac{g^y(Y_{t-1} = y_{t-1} | X_{t-1} = \zeta_{t-1}^i)}{\sum_{j=1}^N g^y(Y_{t-1} = y_{t-1} | X_{t-1} = \zeta_{t-1}^j)}\ \forall i \in \{1, .., N\}$.
  \item Sample $N$ variables $u_i\sim Categorical(w_1, ..., w_N)$.
  \item Sample $N$ particles $\zeta_t^i \sim g(\cdot | X_{t-1} = \zeta_{t-1}^{u_i})$.
  \item Return to step 2, or stop the algorithm if enough time-steps have passed.
\end{enumerate}
\end{algorithm}

Algorithm \ref{algo:particle_filter} describes a typical Particle Filter, where $Z_t = P(Y_{0:t} = y_{0:t})$ and we are interested in extracting $Z_{n-1}$, the likelihood of the observations under our model. We also see that $Z_{n-1}$ is nothing more than the product of the mean of the weights at each time-step, where the weights are the probabilities of observing $Y_t$ if the particles describe the hidden Markov Chain $X_t$. Then at each step, $N$ indexes are sampled, which represent the index of the particles at the previous step, and the ones with the highest weights have the highest probability of being selected. The ones selected are used as the state of the hidden Markov Chain in the previous step and sampled from. Thus, only the particles that fit the data the best are selected and the rest are discarded.

\begin{algorithm}
\caption{Sampling a path}
\begin{enumerate}
\label{algo:sample_path}
  \item Set $t=n-1$, sample a particle from the last time-step at random $I_n \sim U(1, N)$ and set $X_n \leftarrow \zeta_n^{I_n}$.
  \item Follow back the path and set $I_t \leftarrow A_t^{I_{t+1}}$, where $A$ is the ancestral matrix, meaning $A_t^j$ is the index of the ancestor particle of particle $j$ at time $t$.
  \item Set $X_t \leftarrow \zeta_t^{I_t}$.
  \item Iterate $t \leftarrow t-1$ and go to step 2, or end the algorithm if $(X_0, ..., X_n)$ has be sampled.
\end{enumerate}
\end{algorithm}

Then after Algorithm \ref{algo:particle_filter} has been ran, the Algorithm \ref{algo:sample_path} can be run to extract a sample of the hidden Markov Chain $X_t$. The algorithm is pretty straightforward, you select a random particle at the last time-step and then go backwards until you reach time 0 and have a completed chain.

There are various ways we can consider adapting our stochastic SIR model to a Hidden Markov Model. We can consider, for example, having $X_t$ be the hidden $(S_t, I_t, R_t)$ values and $Y_t$ be the $I_t$ value, to simulate the scenario where only infections are observed. Another scenario could be observing only a fraction of the populations of each compartment and model that to retrieve the actual population numbers. Working with hidden Markov Chains makes the models more complicated and computationally expensive, but makes the epidemic models more realistic and, thus, more likely to be used in a real epidemic setting.

\subsection{Particle Markov Chain Monte Carlo}

A class of models used to estimate posterior distribution of random variables is Particle Markov Chain Monte Carlo or Particle-MCMC or PMCMC, which is a combination of Sequential Monte Carlo and an MCMC algorithm. It can be used to estimate the underlying parameters that influence the distributions of a hidden Markov Chain \cite{doucet-2006}. Therefore, a Hidden Markov Model's equations become the following, which is a variation of Equation \ref{eq:hidden1}.

\begin{equation}
\begin{split}
&X_0 \sim \mu\ \textrm{and}\ X_p \sim g(\cdot | X_{p-1}, \theta)\ \textrm{for}\ p\ \textrm{in}\ [1, n] \\
&Y_p \sim g^y(\cdot | X_p, \theta)\ \textrm{for}\ p\ \textrm{in}\ [0, n-1]
\end{split}
\end{equation}

where $\theta$ is unknown.

A PMCMC algorithm estimates the parameter $\theta$ by using an MCMC algorithm to sample $\theta$ and using a Particle Filter to extract the likelihood of that $\theta$. Hence, the Particle Filter is used to extract the likelihoods and a sample hidden Markov Chain at each time-step \cite{andrieu-2010}. The rest of the algorithm is the same MCMC algorithm, as it would have been used on its own. Since we have a stochastic system which evolves through time, PMCMC algorithms take advantage of delaying prediction and using only the best trajectories and, thus, being less computationally intensive.

A common PMCMC algorithm is the particle marginal Metropolis-Hastings algorithm, which is the same as the Metropolis-Hastings algorithm with the addition of using the likelihoods extracted by a Particle Filter to accept a proposed $\theta'$. Define $v$ as the prior distribution of parameter $\theta$, then we have that the posterior distribution of $\theta$ is defined by:

\begin{equation}
\begin{split}
    P(\theta | Y_0, ..., Y_n)
    &\propto v(\theta)P(Y_0, ..., Y_{n-1} | \theta) \\
    &\propto v(\theta) Z_{n,\theta}
\end{split}
\end{equation}

Consequently, the higher the posterior probability of $\theta'$, calculated with likelihood extracted from the particle filter, compared to the posterior probability of the previously accepted $\theta$, the more likely to be accepted.

\begin{algorithm}[H]
\caption{Particle Marginal Metropolis-Hastings Algorithm}
\begin{enumerate}
  \item System is initialised and we have $i=0$, starting time, and sample $\theta_0 \sim v$.
  \item Using the proposal distribution we sample $\theta' \sim h(\cdot | \theta_i)$ and use  $\theta'$ to run the particle filter and extract $Z_{n,\theta'}^N$.
  \item We calculate the probability of acceptance $p_i = \min(1, \frac{v(\theta')Z_{n,\theta'}^N h(\theta_i|\theta')}{v(\theta_i)Z_{n,\theta}^N h(\theta'|\theta_i)})$.
  \item With probability $p_i$ we accept $\theta'$ and $\theta_{i+1} \leftarrow \theta'$ and with probability $1-p_i$ we discard $\theta'$ and $\theta_{i+1} \leftarrow \theta_i$.
  \item Increment $i$ by 1 and return to step 2, until a set number of samples from $\theta$ has been generated.
\end{enumerate}
\end{algorithm}

\section{Results}

In this section I will provide examples of the uses of the algorithms mentioned in the "Methodology" Section \ref{sec:meth} in an epidemiological setting. The models were fitted using simulated data, with known parameters, to validate their performance.

The code used to run the algorithms can be found here: \url{https://github.com/GeorgeEfstathiadis/Stochastic-Epidemic-Modelling}

\subsection{Noisy SIR model} \label{sec:noisy_sir}

I simulate a deterministic SIR epidemic with parameters $\beta = 2$ and $\gamma = 1$. The total population is 4820 with 4800 people in the susceptible compartment and 20 in the infectious compartment. The epidemic is simulated in continuous time, but only keeping the values at integer time-points. I then proceed to add some Normal noise to imitate real world settings, where the disease is not following the exact deterministic SIR model. The noise is independent between different compartments and time-points and is described by $e_p \sim N_3(0, n_{ratio} * X_p * \mathbf{I}_3)$, where $p$ represents the time, $X_p$ the simulated, hidden, data at time $p$ and $n_{ratio}$ is a predetermined ratio to enhance the noise depending on the size of the compartment at a given time. For this case I selected 0.01 as our $n_{ratio}$. Thus the distribution $g^y(\cdot | X_p)$ of the Hidden Markov Chain is $N_3(X_p, 0.01 * X_p * \mathbf{I}_3)$.

I proceed to model this simulated dataset using 2 different algorithms, ABC and PMCMC.

\subsubsection{ABC}

I assume $\beta$ and $\gamma$ have independent uniform priors in the range of (0, 5). For function $g$ to simulate $X_i$ the Gillespie algorithm is used and for summary statistic function $S$ the infected and removed counts at each time-point are used. Finally, for distance function the mean absolute difference function is used and for threshold $\epsilon$, after a trial process, I ended up with 150.

I gathered 1,000 samples to form the posterior distribution and the results were as follows.

\begin{figure}[H]
\caption{Density plot of posterior samples of $\beta$ and $\gamma$ from ABC.}
\centering
\includegraphics[width=0.7\textwidth]{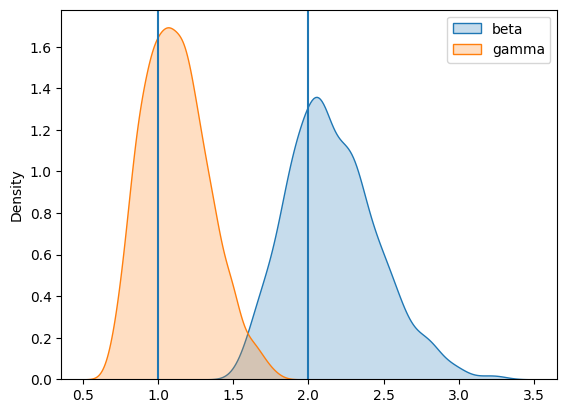}
\end{figure}

\begin{figure}[H]
\caption{Sampled trajectories from ABC.}
\centering
\includegraphics[width=0.7\textwidth]{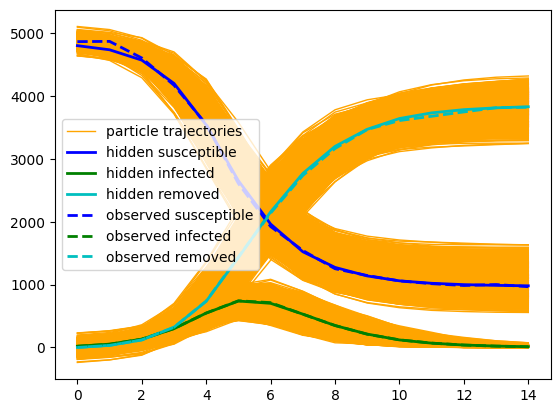}
\end{figure}

\begin{figure}[H]
\caption{Sampled 95\% credible intervals from ABC.}
\centering
\includegraphics[width=0.7\textwidth]{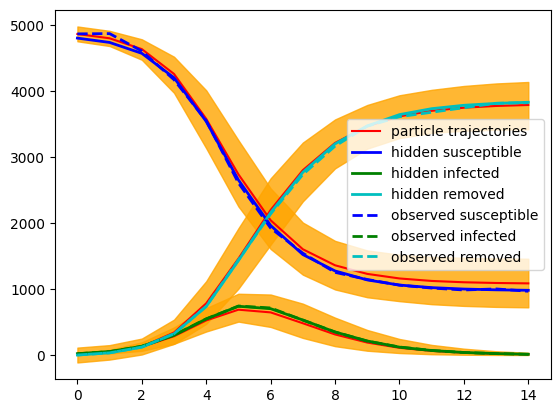}
\end{figure}

Overall, the use of a non informative prior increased the computation time and we can clearly observe that it influenced the posterior distribution more than we would have wanted. For $\beta$ the posterior mean is 2.16 with a $95\%$ highest posterior density interval of (1.626, 2.768), while for $\gamma$ the posterior mean is 1.12 with a $95\%$ highest posterior density interval of (0.729, 1.52). The posterior median value of $\beta$ is 2.12 and of $\gamma$ is 1.10.

The algorithm achieves a decent performance, but this can be improved by using a more informative prior and a more complicated algorithm. In this setting ABC will not be able to scale well with more complicated scenarios.

\subsubsection{PMCMC} \label{sec:pmcmc_noisy}

I also model the noisy SIR dataset using a PMCMC algorithm. For the prior and proposal distribution of $\beta$ and $\gamma$ I use a multivariate normal distribution \cite{s-rosenthal-2011}. In consequence, the distribution looks similar to this:

\begin{equation}
(\beta_{i+1}, \gamma_{i+1}) \sim N((\beta_i, \gamma_i), h\pmb{\Sigma})
\end{equation}

where $\pmb{\Sigma}$ is the estimated posterior covariance matrix of $\beta$ and $\gamma$ and $h$ is a constant that adjusts the covariance of the proposal distribution so that we have an acceptance rate between $10\%$ and $25\%$. I also add a constraint to make sure our parameters are strictly larger than 0, while if any of the proposed parameters are below 0 we reject that sample.

I run the algorithm twice for around 1,000 epochs, one for estimating $\pmb{\Sigma}$ and one for estimating $h$, without saving the results to the posterior distribution, a variation to the adaptive approach Haario and others formulated \cite{haario-2001}. Hence, when I run the final algorithm, I have already estimated $\pmb{\Sigma}$, $h$ and the prior distribution has as its mean the final sample of $\beta$ and $\gamma$ from the previous test run.

I then fix $\pmb{\Sigma}$ and $h$ and run the algorithm. I run the algorithm on 3 different chains to ensure convergence for 25,000 steps each. Since the process can be quite computationally intensive, I make sure to use parallel processing to receive our results faster. 
I also run tests on the number of particles needed to have consistent likelihoods returned from the particle filter for the same parameters. I noticed that difference in variance between the likelihoods for 10 and 100 particles was quite significant, but between 100 and 1000 was not significant. Thus I decided to run PMCMC with 100 particles in the interest of reducing the computational time.

After receiving the posterior samples of $\beta$ and $\gamma$, I apply a burn-in of 1,000 samples, meaning I removed the first 1,000 samples received, in order to ensure that the parameters have reached convergence, before adding them to the posterior distribution. I also applied a thinning of 40 samples, meaning I only kept 1 sample every 40 samples for the posterior distribution, to reduce the auto-correlation between consecutive samples. After that I receive the estimated posterior distribution presented below. I also perform the Gelman-Rubin test to ensure all 3 chains have converged, which comes up with 1.00030734 and 1.00002084 for $\beta$ and $\gamma$ respectively, which are indeed sufficiently close to 1. I also calculate the effective sample size to see how many samples are independent by accounting for auto-correlation \cite{kish-1965}, which ends up being 1267 and 861 for each parameter.

\begin{figure}[H]
\caption{Posterior samples of $\beta$ from PMCMC on noisy data.}
\centering
\includegraphics[width=0.7\textwidth]{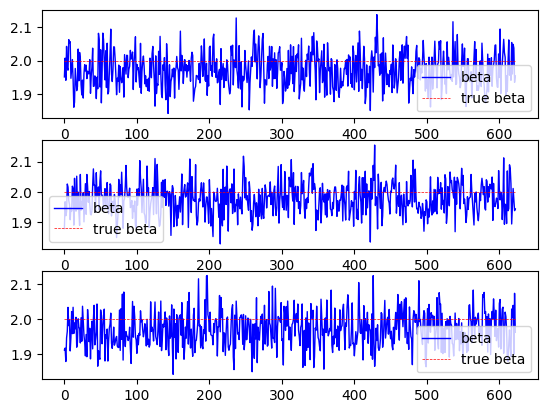}
\end{figure}

\begin{figure}[H]
\caption{Posterior samples of $\gamma$ from PMCMC on noisy data.}
\centering
\includegraphics[width=0.7\textwidth]{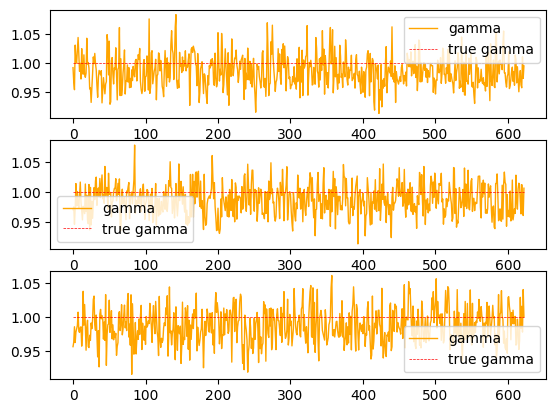}
\end{figure}

\begin{figure}[H]
\caption{Density plot of posterior samples of $\beta$ and $\gamma$ from PMCMC on noisy data.}
\centering
\includegraphics[width=0.7\textwidth]{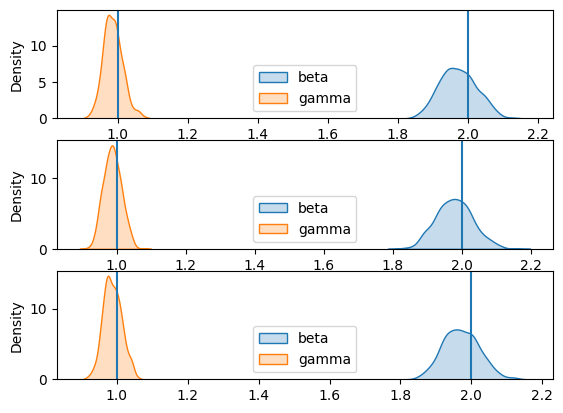}
\end{figure}

\begin{figure}[H]
\caption{Line plot of likelihoods times prior for each sample from PMCMC on noisy data.}
\centering
\includegraphics[width=0.7\textwidth]{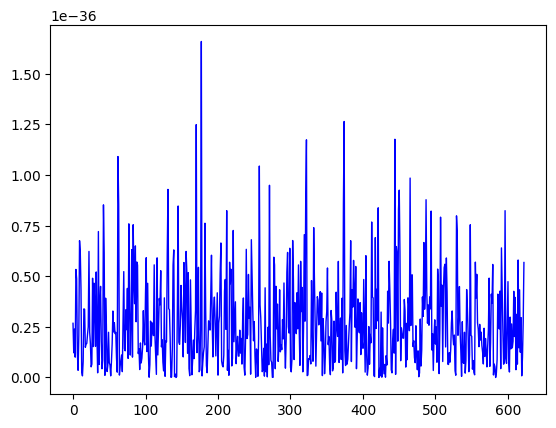}
\end{figure}

\begin{figure}[H]
\caption{Sampled trajectories from PMCMC on noisy data.}
\centering
\includegraphics[width=0.7\textwidth]{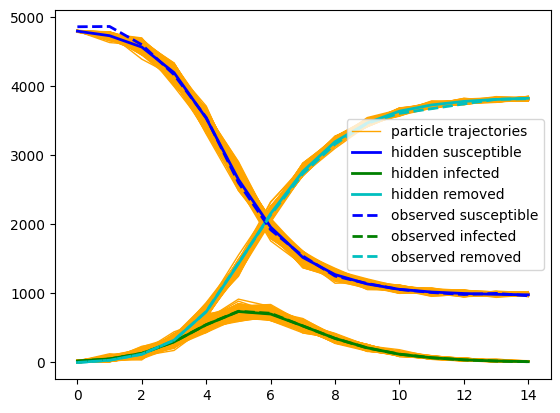}
\end{figure}

\begin{figure}[H]
\caption{Sampled 95\% credible intervals from PMCMC on noisy data.}
\centering
\includegraphics[width=0.7\textwidth]{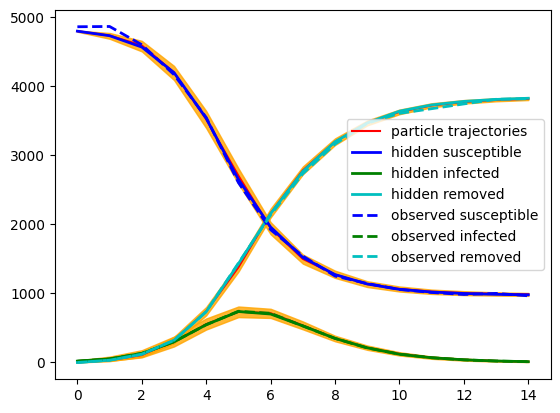}
\end{figure}

The acceptance rate, before applying thinning is around 22\%, while after applying thinning is around 100\%. We also have the estimated posterior covariance is equal to:

\begin{equation}
    \pmb{\Sigma} = \begin{pmatrix}
        0.00269008 & 0.0007161\\
        0.0007161 & 0.00075565
    \end{pmatrix}
\end{equation}

The posterior mean of $\beta$ is 1.973 with a $95\%$ highest posterior density interval of (1.884, 2.079), while for $\gamma$ the posterior mean is 0.986 with a $95\%$ highest posterior density interval of (0.927, 1.035). The posterior median of $\beta$ is 1.972 and of $\gamma$ is 0.984.

\subsection{Under-reported SIR model}

I use the same simulated dataset from the previous Subsection \ref{sec:noisy_sir}, with $\beta$ equal to 2 and $\gamma$ equal to 1 and with a population of 4820 with 4800 people in the susceptible compartment and 20 in the infectious compartment. Instead of adding Normal noise I only use a subset of our data to imitate real world under-reported data. I generate our observed data by applying a Binomial distribution to the, real, hidden data with a probability of observation being $p_{obs}$. Thus, the distribution $g^y(\cdot | X_p)$ of the Hidden Markov Chain is 3 independent $Bin(X_p, \ p_{obs})$ distributions, one for each compartment.

I proceed by modelling two different scenarios that might be useful in a real world epidemic. In the first model I assume that we know the probability of observing an infection and on the second scenario I assume that we don't know it and try to estimate the posterior of the probability of observation as well. For both scenarios I am using the PMCMC model and our observed data are generated using $p_{obs}$ equal to 0.1.

\subsubsection{PMCMC with known missing rate}

Similarly to the scenario discussed previously on noisy data, I use a multivariate normal distribution for the prior and proposal distributions of $\beta$ and $\gamma$. I run two chains with around 1,000 epochs, one for estimating $\pmb{\Sigma}$ and one for estimating $h$, without saving the results and choose an $h$ and $\pmb{\Sigma}$ than gives us a desirable acceptance rate as done in Subsection \ref{sec:pmcmc_noisy}.

After some analysis of the likelihoods I decide to use 1,000 particles and run the algorithm on 3 different chains with 25,000 steps each to ensure convergence. When the algorithm ends, I apply a burn-in of 1,000 samples and a thinning of 40. I also perform the Gelman-Rubin test diagnostic to ensure convergence on the 3 chains and I get 1.00030734 and 1.00002084 statistics for $\beta$ and $\gamma$ respectively, which are sufficiently close to 1. I also calculate the effective sample size, which ends up being 1050, 1300 for each parameter.

\begin{figure}[H]
\caption{Posterior samples of $\beta$ from PMCMC on under-reported data with known missing rate.}
\centering
\includegraphics[width=0.7\textwidth]{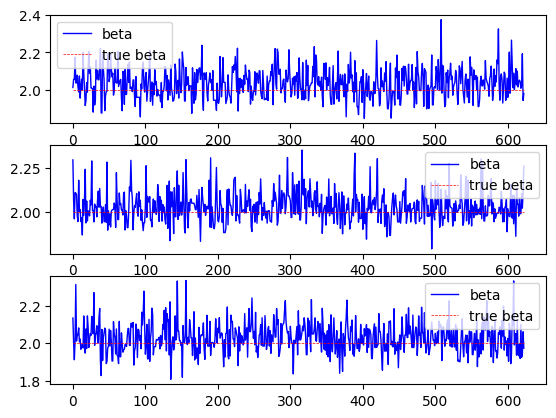}
\end{figure}

\begin{figure}[H]
\caption{Posterior samples of $\gamma$ from PMCMC on under-reported data with known missing rate.}
\centering
\includegraphics[width=0.7\textwidth]{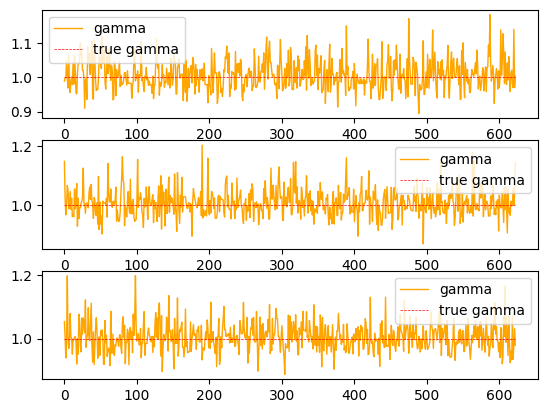}
\end{figure}

\begin{figure}[H]
\caption{Density plot of posterior samples of $\beta$ and $\gamma$ from PMCMC on under-reported data with known missing rate.}
\centering
\includegraphics[width=0.7\textwidth]{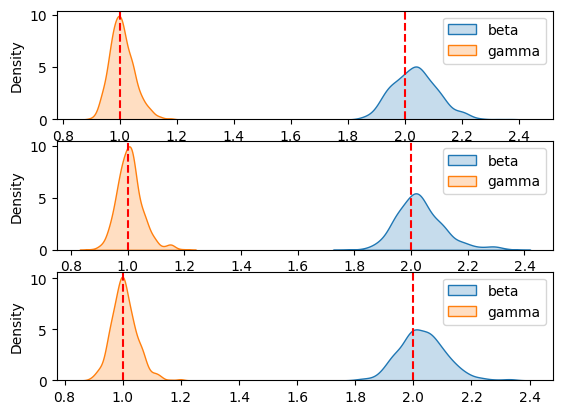}
\end{figure}

\begin{figure}[H]
\caption{Line plot of likelihoods times prior for each sample from PMCMC on under-reported data with known missing rate.}
\centering
\includegraphics[width=0.7\textwidth]{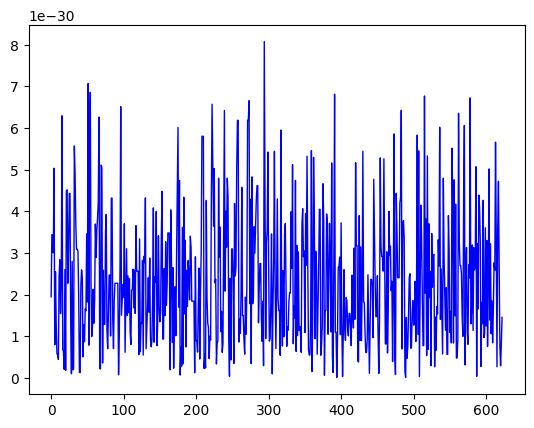}
\end{figure}

\begin{figure}[H]
\caption{Sampled trajectories from PMCMC on under-reported data with known missing rate.}
\centering
\includegraphics[width=0.7\textwidth]{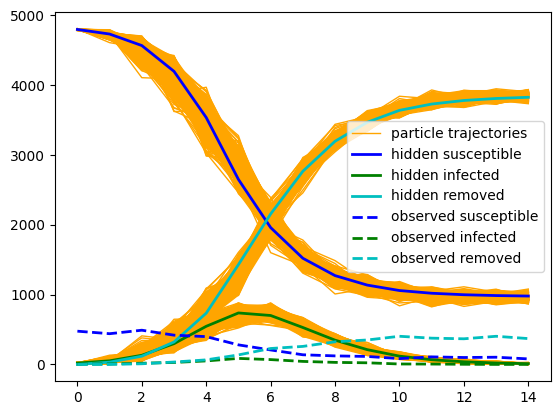}
\end{figure}

\begin{figure}[H]
\caption{Sampled 95\% credible intervals from PMCMC on under-reported data with known missing rate.}
\centering
\includegraphics[width=0.7\textwidth]{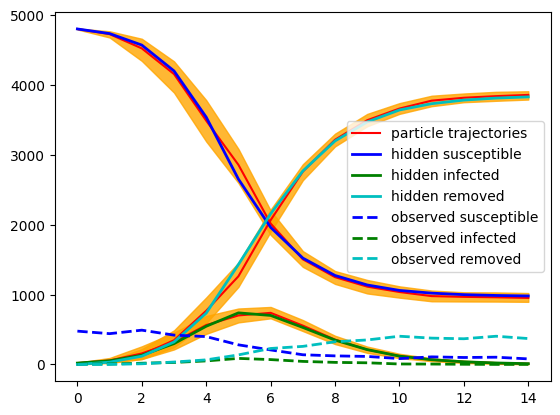}
\end{figure}

Before applying the thinning we have an acceptance rate of around 12\% and after applying thinning we get an acceptance rate of around 95\%. We also have the estimated posterior covariance being:

\begin{equation}
    \pmb{\Sigma} = \begin{pmatrix}
        0.00653857 & 0.00305867\\
        0.00305867 & 0.0019865
    \end{pmatrix}
\end{equation}

For $\beta$ a posterior mean of around 2.036 is achieved with a 95\% highest posterior density interval of (1.907, 2.205), while for $\gamma$ a posterior mean of 1.008 is achieved with a 95\% highest posterior density interval of (0.93, 1.094). The posterior median of $\beta$ is around 2.036 and the posterior median of $\gamma$ is around 1.003.

\subsubsection{PMCMC with unknown missing rate}

Modelling the scenario with an unknown missing rate has one difference from the previous scenario; there is one extra variable's posterior distribution to estimate. I follow the same steps as in the previous subsections but the prior and proposal distributions are now 3-dimensional since they include $p_{obs}$ as the third variable.

After some analysis of the likelihoods I decide to use 100 particles again and run the algorithm on 3 different chains with 25,000 steps each to ensure convergence. When the algorithm ends, I apply a burn-in of 1,000 samples and a thinning of 40. I also perform the Gelman-Rubin test to ensure convergence and end up with 1.00040369, 1.00070195 and 1.00025463 for $\beta$, $\gamma$ and $p_{obs}$ respectively, which are all sufficiently close to 1. I also calculate the effective sample size, which ends up being 1275, 1732 and 2622 for each parameter.

\begin{figure}[H]
\caption{Posterior samples of $\beta$ from PMCMC on under-reported data with unknown missing rate.}
\centering
\includegraphics[width=0.7\textwidth]{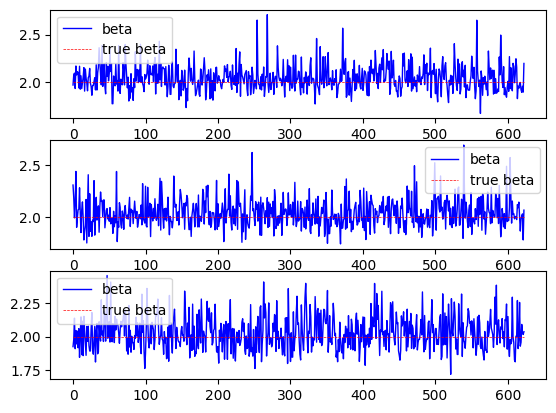}
\end{figure}

\begin{figure}[H]
\caption{Posterior samples of $\gamma$ from PMCMC on under-reported data with unknown missing rate.}
\centering
\includegraphics[width=0.7\textwidth]{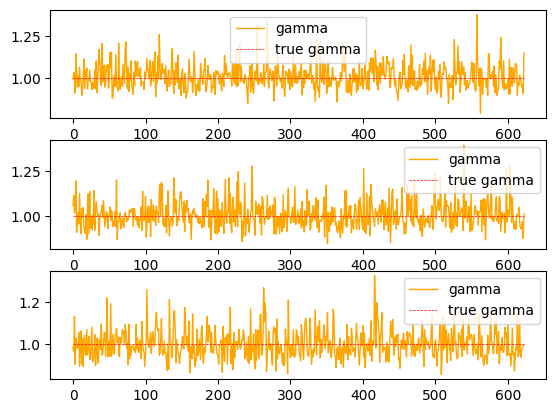}
\end{figure}

\begin{figure}[H]
\caption{Posterior samples of $p_{obs}$ from PMCMC on under-reported data with unknown missing rate.}
\centering
\includegraphics[width=0.7\textwidth]{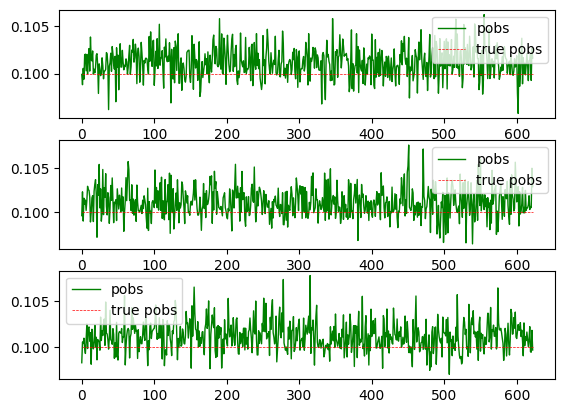}
\end{figure}

\begin{figure}[H]
\caption{Density plot of posterior samples of $\beta$ and $\gamma$ from PMCMC on under-reported data with unknown missing rate.}
\centering
\includegraphics[width=0.7\textwidth]{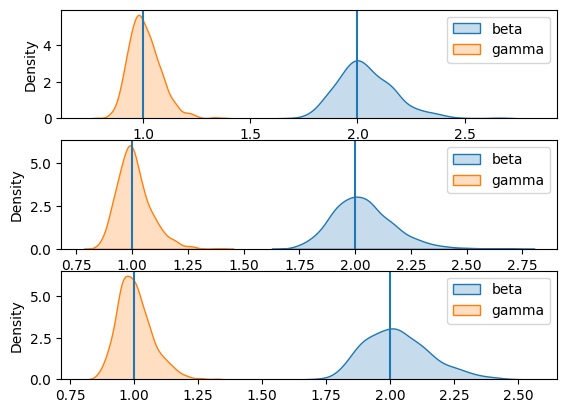}
\end{figure}

\begin{figure}[H]
\caption{Density plot of posterior samples of $p_{obs}$ from PMCMC on under-reported data with unknown missing rate.}
\centering
\includegraphics[width=0.7\textwidth]{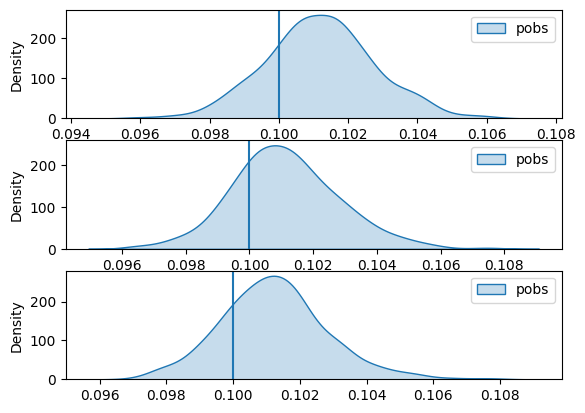}
\end{figure}

\begin{figure}[H]
\caption{Line plot of likelihoods times prior for each sample from PMCMC on under-reported data with unknown missing rate.}
\centering
\includegraphics[width=0.7\textwidth]{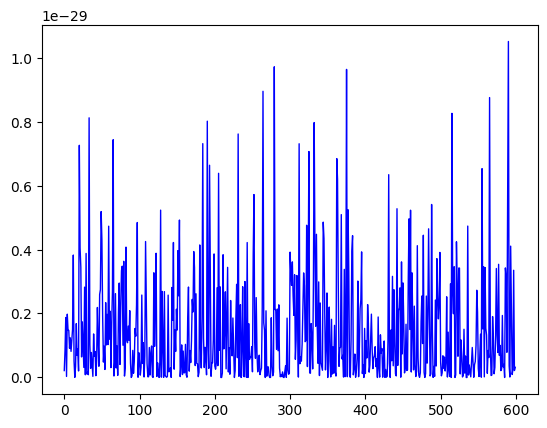}
\end{figure}

\begin{figure}[H]
\caption{Sampled trajectories from PMCMC on under-reported data with unknown missing rate.}
\centering
\includegraphics[width=0.7\textwidth]{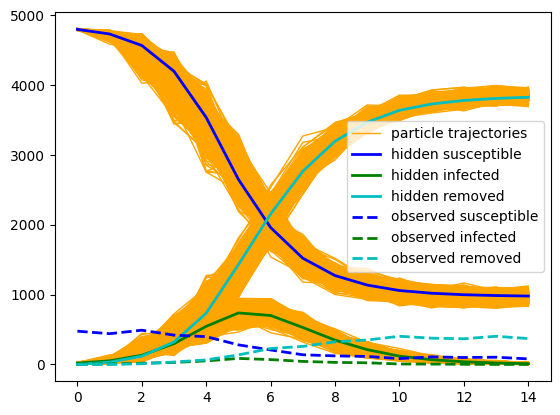}
\end{figure}

\begin{figure}[H]
\caption{Sampled 95\% credible intervals from PMCMC on under-reported data with unknown missing rate.}
\centering
\includegraphics[width=0.7\textwidth]{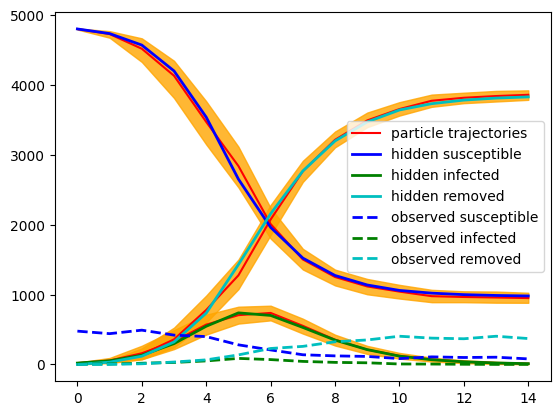}
\end{figure}

Before thinning the acceptance rate is around 23\%, while after applying thinning the acceptance rate is around 86\%. Meanwhile, the estimated posterior covariance is equal to:

\begin{equation}
    \pmb{\Sigma} = \begin{pmatrix}
        1.88909760e-02 & 9.22455177e-03 & -3.72354364e-05\\
        9.22455177e-03 & 5.86458070e-03 & -2.10057520e-05\\
        -3.72354364e-05 & -2.10057520e-05 & 3.03817595e-06
    \end{pmatrix}
\end{equation}

The posterior mean of $\beta$ is 2.034 with a 95\% highest posterior density interval of (1.79, 2.311), the posterior mean of $\gamma$ is 1.006 with a 95\% highest posterior density interval of (0.877, 1.16) and the posterior mean of $p_{obs}$ is 0.101 with a 95\% highest posterior density interval of (0.098, 0.105). The posterior median of $\beta$ is 2.018, of $\gamma$ is 0.977 and of $p_{obs}$ is 0.101.

\subsection{Under-reported SEIR model}

I am now going to use a slightly more complicated model than the SIR, which was discussed in Subsection \ref{sec:compartment}, the SEIR model. This model includes an extra parameter, $\alpha$, which signifies the rate at which people from the susceptible group enter the exposed group. I will work with under-reported data again, generated by applying the binomial distribution to the simulated data, but I am only going to model the scenario where the probability of observation is known. This is done as proof of concept and expanding it to the scenario where $p_{obs}$ is unknown would be modelled the same way.

I generate our data from a deterministic SEIR model with $\beta = 4$, $\gamma = 1$ and $\alpha = 1$. Total population is 4820 people, with 4800 people in the susceptible group, 20 in the infected group and 0 in both the exposed and removed group. I then apply a binomial distribution with $p_{obs} = 0.1$ to imitate real world conditions, when data are under-reported. Thus, the distribution of the hidden Markov Chain is the same as in the previous section.

I model this scenario using the PMCMC algorithm. For the prior and proposal distribution of $\beta$, $\gamma$ and $\alpha$ I use a multivariate normal distribution, as in the previous subsections, which is 3-dimensional. I run the algorithm with 3 chains using 100 particles and for 25,000 steps each. After the algorithm is finished I apply a burn-in of 1,000 samples and thinning of 20. I also perform the Gelman-Rubin test to ensure convergence, which ends up being 1.00103306, 1.00010529 and 1.00098897 for $\beta$, $\gamma$ and $\alpha$ respectively, which are sufficiently close to 1. I also calculate the effective sample size, which ends up being 752, 2065 and 835 for each parameter.

\begin{figure}[H]
\caption{Posterior samples of $\beta$ from PMCMC on SEIR under-reported data.}
\centering
\includegraphics[width=0.7\textwidth]{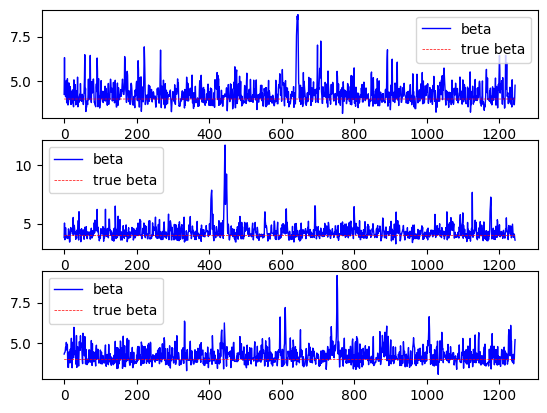}
\end{figure}

\begin{figure}[H]
\caption{Posterior samples of $\gamma$ from PMCMC on SEIR under-reported data.}
\centering
\includegraphics[width=0.7\textwidth]{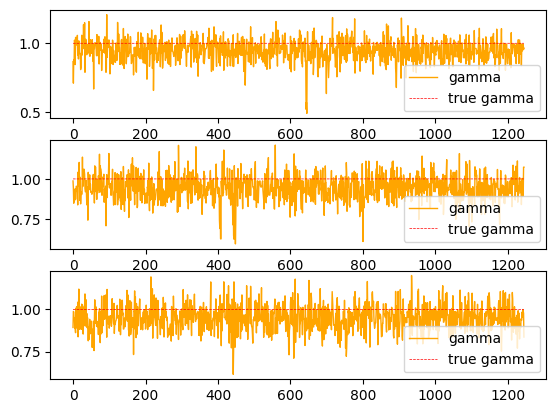}
\end{figure}

\begin{figure}[H]
\caption{Posterior samples of $\alpha$ from PMCMC on SEIR under-reported data.}
\centering
\includegraphics[width=0.7\textwidth]{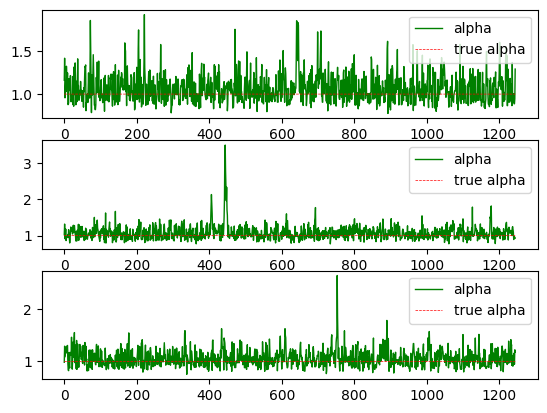}
\end{figure}

\begin{figure}[H]
\caption{Density plot of posterior samples of $\beta$ from PMCMC on SEIR under-reported data.}
\centering
\includegraphics[width=0.7\textwidth]{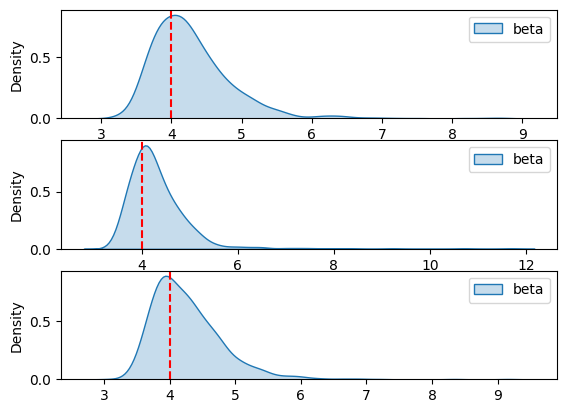}
\end{figure}

\begin{figure}[H]
\caption{Density plot of posterior samples of $\gamma$ from PMCMC on SEIR under-reported data.}
\centering
\includegraphics[width=0.7\textwidth]{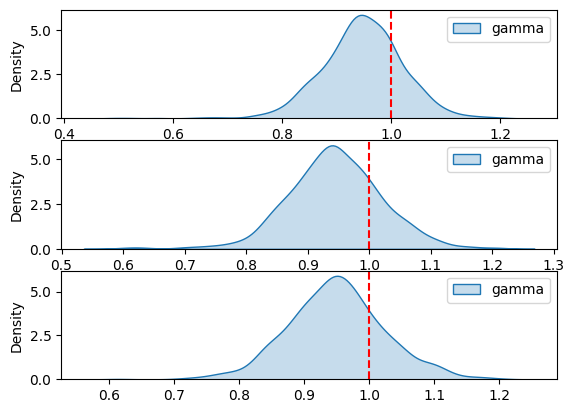}
\end{figure}

\begin{figure}[H]
\caption{Density plot of posterior samples of $\alpha$ from PMCMC on SEIR under-reported data.}
\centering
\includegraphics[width=0.7\textwidth]{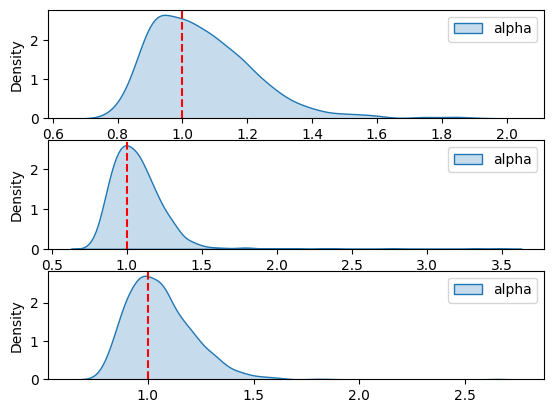}
\end{figure}

\begin{figure}[H]
\caption{Line plot of likelihoods times prior for each sample from PMCMC on SEIR under-reported data.}
\centering
\includegraphics[width=0.7\textwidth]{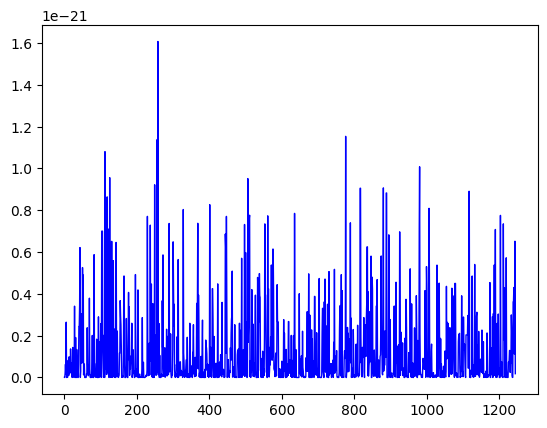}
\end{figure}

\begin{figure}[H]
\caption{Sampled trajectories from PMCMC on SEIR under-reported data.}
\centering
\includegraphics[width=0.7\textwidth]{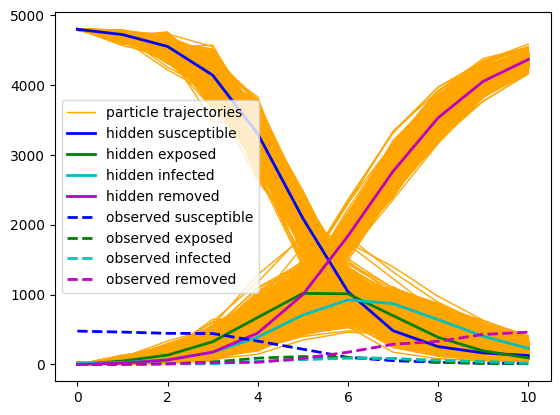}
\end{figure}

\begin{figure}[H]
\caption{Sampled 95\% credible intervals from PMCMC on SEIR under-reported data.}
\centering
\includegraphics[width=0.7\textwidth]{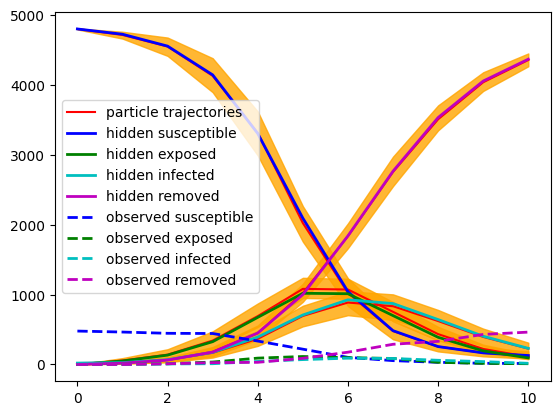}
\end{figure}

Before thinning the acceptance rate is around 19.7\%, while after applying thinning the acceptance rate is around 93.3\%. We also have the estimated posterior covariance being equal to:

\begin{equation}
    \Sigma = \begin{pmatrix}
        0.38062885 & -0.02176196 & 0.08728822\\
        -0.02176196 & 0.00606955 & -0.00338927\\
        0.08728822 & -0.00338927 &  0.02770316
    \end{pmatrix}
\end{equation}

For $\beta$ a posterior mean of 4.294 is achieved with a 95\% highest posterior density interval of (3.462, 5.443), for $\gamma$ a posterior mean of 0.946 is achieved with a 95\% highest posterior density interval of (0.803, 1.091) and for $\alpha$ a posterior mean of 1.064 is achieved with a 95\% highest posterior density interval of (0.797, 1.362). The posterior medians are 4.178, 0.948 and 1.034 for $\beta$, $\gamma$ and $\alpha$ respectively.

\subsection{Experiments on information reduction}

The following section examines various experiments that were performed on reducing the amount of data the algorithms are trained on to pinpoint the effects of these changes on convergence. It is of interest to find out how much the observed datasets can be modified, with the chains still converging to their posterior densities.

In order to assess the performance of the algorithm in the modified datasets I used the density plots of the posterior density and the trajectory plots. I also calculated the posterior means and high posterior density intervals of the underlying parameters, as well as their posterior mean squared error defined by:

\begin{equation}
    PMSE_i = \frac{\sum_{j=1}^n(\theta_i-\theta'_{i,j})^2}{n}
\end{equation}

where $\theta_i$ is the $i^{th}$ true parameter used and $\theta'_{i,j}$ it the $j^{th}$ posterior sample of the $i^{th}$ parameter from the estimated posterior samples.

For the purposes of this experimentation on the amount of information needed I am using the adaptive version of the algorithm used in the previous subsections proposed by Haario and others \cite{haario-2001}. In essence, I fix $h$ at a sensible value, using prior knowledge, and I adapt the posterior covariance of the parameters $\Sigma$, while the algorithm runs. Thus, the formula for $\Sigma$ now is:

\begin{equation}
    \bm{\Sigma}_t = 
        \begin{cases}
            \bm{I}_n &\quad\text{if }t \leq t_0 \\
            \text{cov}(\bm{\theta}_0, ..., \bm{\theta}_{t-1}) + \epsilon \bm{I}_n &\quad\text{if }t > t_0
        \end{cases}
\end{equation}

where $\bm{I}_n$ is the identity matrix with $n$ dimensions, $n$ being the dimensions of $\theta$. $\epsilon$ is fixed to be 0.0001 and $t_0$ to be 1,000. Thus, we do not start adapting the posterior covariance until the first 1,000 steps have passed and then applying a burn-in of 1,000 steps to each chain.

Hence, the new proposal distribution inside the PMCMC will be $\theta_{t+1} \sim N_n(\theta_t, h\pmb{\Sigma}_t)$. The difference with the results we got from the previous subsections is that now the results are no longer Markov Chains, since a sample $\theta_{t+1}$ will depend on all previous samples, due to the adaptive nature of the posterior covariance.

\subsubsection{Noise level} \label{sec:noise_level}

The results of increasing the noise, for the same SIR noisy model discussed above, are investigated here. Originally, convergence was achieved on the dataset with a noise level of 0.01. For this experiment, I ran the algorithm on datasets created using a $n_{ratio}$ of 0.05, 0.1, 0.15, 0.2 and 0.25. I run one chain for each $n_{ratio}$ used, with 6,000 samples each with a burn-in of 1,000 samples and a thinning of 10 samples for the visualisations only.

For $n_{ratio}$ equal to 0.05, the posterior mean of $\beta$ was 2.018 with a high posterior density interval of (1.917, 2.142), while the posterior mean of $\gamma$ was 1.009 with a high posterior density interval of (0.954, 1.068). The posterior mean squared error of $\beta$ and $\gamma$ were 0.004008 and 0.00097 respectively.

\begin{figure}[H]
\caption{Density plot of posterior samples of $\beta$ and $\gamma$ from PMCMC on SIR noisy data with $n_{ratio}$=0.05.}
\centering
\includegraphics[width=0.7\textwidth]{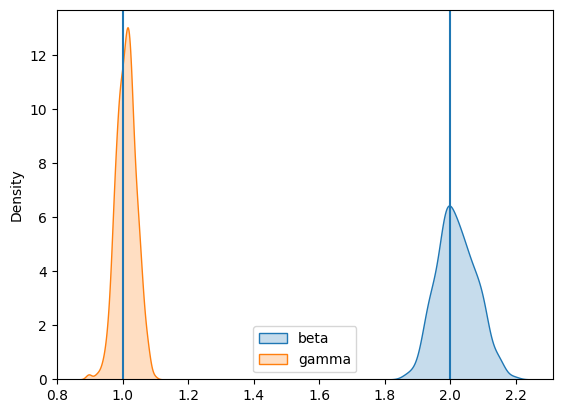}
\end{figure}

\begin{figure}[H]
\caption{Sampled 95\% credible intervals from PMCMC on SIR noisy data with $n_{ratio}$=0.05.}
\centering
\includegraphics[width=0.7\textwidth]{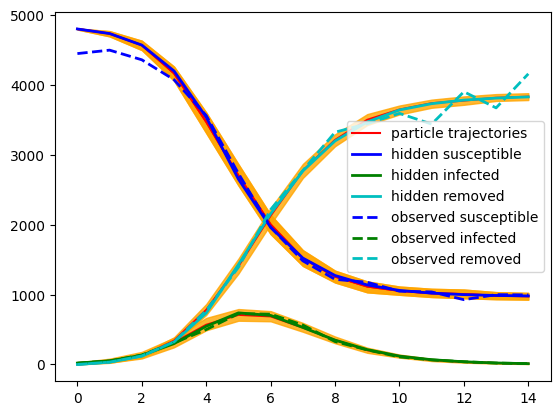}
\end{figure}

For $n_{ratio}$ equal to 0.1, the posterior mean of $\beta$ was 1.944 with a high posterior density interval of (1.796, 2.097), while the posterior mean of $\gamma$ was 0.956 with a high posterior density interval of (0.874, 1.031). The posterior mean squared error of $\beta$ and $\gamma$ were 0.009579 and 0.003718 respectively.

\begin{figure}[H]
\caption{Density plot of posterior samples of $\beta$ and $\gamma$ from PMCMC on SIR noisy data with $n_{ratio}$=0.1.}
\centering
\includegraphics[width=0.7\textwidth]{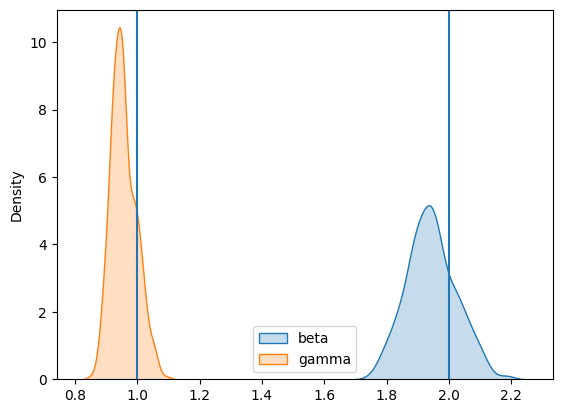}
\end{figure}

\begin{figure}[H]
\caption{Sampled 95\% credible intervals from PMCMC on SIR noisy data with $n_{ratio}$=0.1.}
\centering
\includegraphics[width=0.7\textwidth]{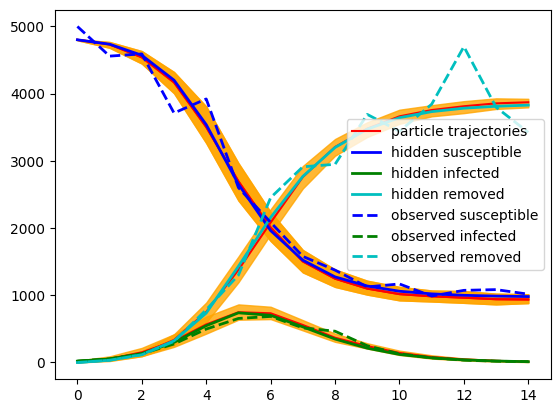}
\end{figure}

For $n_{ratio}$ equal to 0.15, the posterior mean of $\beta$ was 1.999 with a high posterior density interval of (1.79, 2.197), while the posterior mean of $\gamma$ was 0.954 with a high posterior density interval of (0.833, 1.066). The posterior mean squared error of $\beta$ and $\gamma$ were 0.011788 and 0.005968 respectively.

\begin{figure}[H]
\caption{Density plot of posterior samples of $\beta$ and $\gamma$ from PMCMC on SIR noisy data with $n_{ratio}$=0.15.}
\centering
\includegraphics[width=0.7\textwidth]{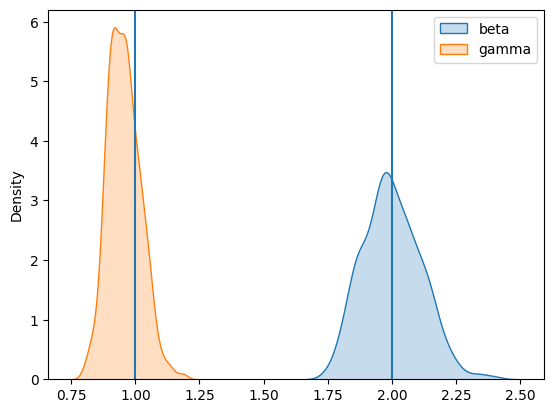}
\end{figure}

\begin{figure}[H]
\caption{Sampled 95\% credible intervals from PMCMC on SIR noisy data with $n_{ratio}$=0.15.}
\centering
\includegraphics[width=0.7\textwidth]{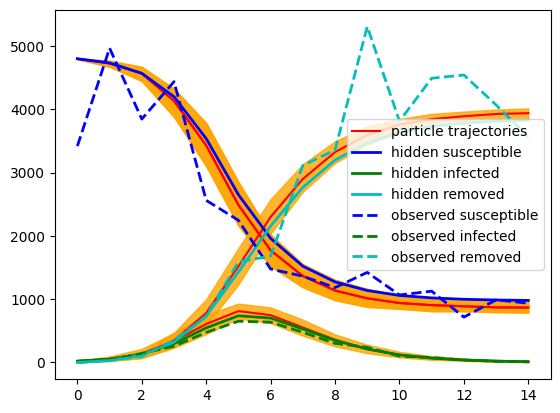}
\end{figure}

For $n_{ratio}$ equal to 0.2, the posterior mean of $\beta$ was 1.907 with a high posterior density interval of (1.688, 2.146), while the posterior mean of $\gamma$ was 0.94 with a high posterior density interval of (0.815, 1.071). The posterior mean squared error of $\beta$ and $\gamma$ were 0.022985 and 0.008295 respectively.

\begin{figure}[H]
\caption{Density plot of posterior samples of $\beta$ and $\gamma$ from PMCMC on SIR noisy data with $n_{ratio}$=0.2.}
\centering
\includegraphics[width=0.7\textwidth]{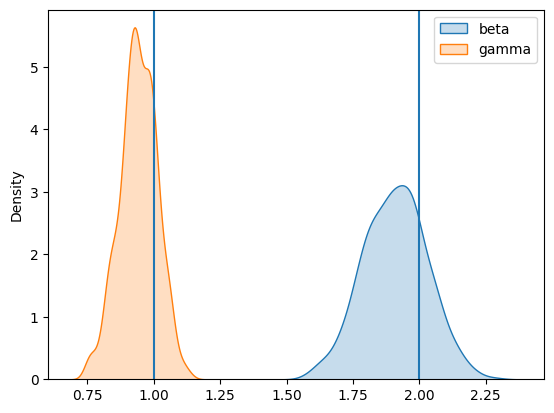}
\end{figure}

\begin{figure}[H]
\caption{Sampled 95\% credible intervals from PMCMC on SIR noisy data with $n_{ratio}$=0.2.}
\centering
\includegraphics[width=0.7\textwidth]{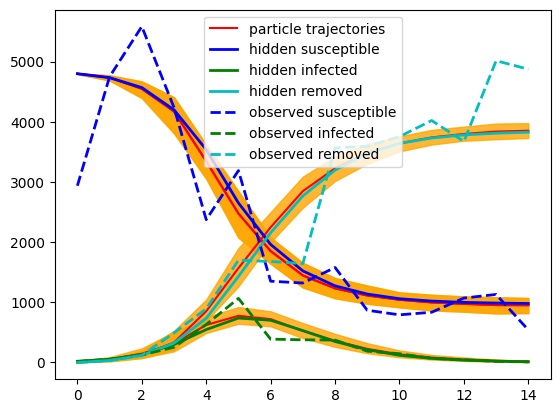}
\end{figure}

For $n_{ratio}$ equal to 0.25, the posterior mean of $\beta$ was 1.96 with a high posterior density interval of (1.685, 2.268), while the posterior mean of $\gamma$ was 0.992 with a high posterior density interval of (0.827, 1.175). The posterior mean squared error of $\beta$ and $\gamma$ were 0.02266 and 0.008478 respectively.

\begin{figure}[H]
\caption{Density plot of posterior samples of $\beta$ and $\gamma$ from PMCMC on SIR noisy data with $n_{ratio}$=0.25.}
\centering
\includegraphics[width=0.7\textwidth]{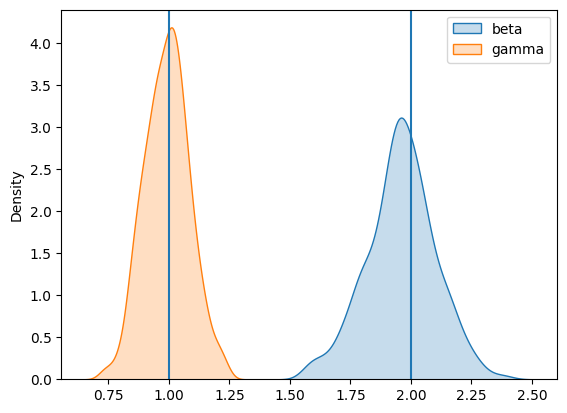}
\end{figure}

\begin{figure}[H]
\caption{Sampled 95\% credible intervals from PMCMC on SIR noisy data with $n_{ratio}$=0.25.}
\centering
\includegraphics[width=0.7\textwidth]{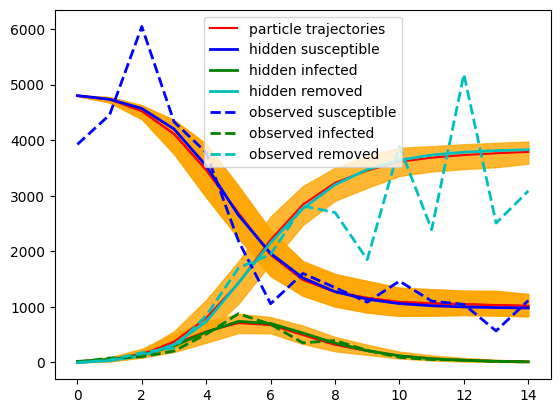}
\end{figure}

For all the different noise levels the chains were able to produce fairly good results. While the posterior means remain close to the true value and the high posterior density intervals still include the true values, we can observe that as the noise level increases, so does the posterior variance of the parameters. We can see that visually from the density plots and the credible interval of the trajectories, which get wider. We can also deduce that from the posterior mean squared error increasing. The posterior bias squared remains relatively stable, thus the posterior variance is causing the increase. This is to be expected, since as the noise increases the more unsure we are about the posterior estimates. Thus, while increasing the noise, the posterior estimates are still very close to the true values, the posterior mean squared error remains relatively low and the sampled trajectories are still close to the truth.

\subsubsection{Probability of observation level}

In the following subsection the results on experimenting with the probability of observation, $p_{obs}$, for the under-reported SIR model with known $p_{obs}$ are presented. In the previous subsection the algorithm was fitted for a $p_{obs}$ equal to 0.1. Now, I will investigate what happens if $p_{obs}$ is equal to 0.075, 0.05, 0.025 and 0.01. One chain is run for each $p_{obs}$ used, with 6,000 samples each with a burn-in of 1,000 samples and a thinning of 10 samples for the visualisations only.

For the 0.075 $p_{obs}$, the posterior mean of $\beta$ was 2.013 with a high posterior density interval of (1.84, 2.217), while the posterior mean of $\gamma$ was 1.0 with a high posterior density interval of (0.883, 1.104). The posterior mean squared error of $\beta$ and $\gamma$ were 0.009624 and 0.003027 respectively.

\begin{figure}[H]
\caption{Density plot of posterior samples of $\beta$ and $\gamma$ from PMCMC on SIR under-reported data with $p_{obs}$=0.075.}
\centering
\includegraphics[width=0.7\textwidth]{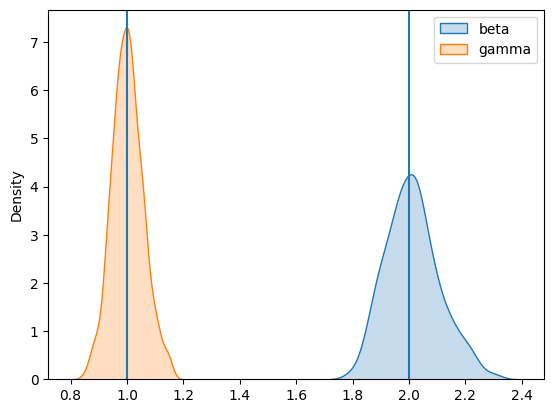}
\end{figure}

\begin{figure}[H]
\caption{Sampled 95\% credible intervals from PMCMC on SIR under-reported data with $p_{obs}$=0.075.}
\centering
\includegraphics[width=0.7\textwidth]{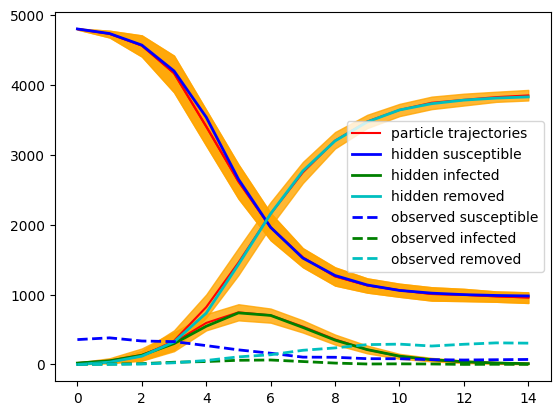}
\end{figure}

For the 0.05 $p_{obs}$, the posterior mean of $\beta$ was 2.074 with a high posterior density interval of (1.838, 2.355), while the posterior mean of $\gamma$ was 1.067 with a high posterior density interval of (0.925, 1.221). The posterior mean squared error of $\beta$ and $\gamma$ were 0.024057 and 0.010465 respectively.

\begin{figure}[H]
\caption{Density plot of posterior samples of $\beta$ and $\gamma$ from PMCMC on SIR under-reported data with $p_{obs}$=0.05.}
\centering
\includegraphics[width=0.7\textwidth]{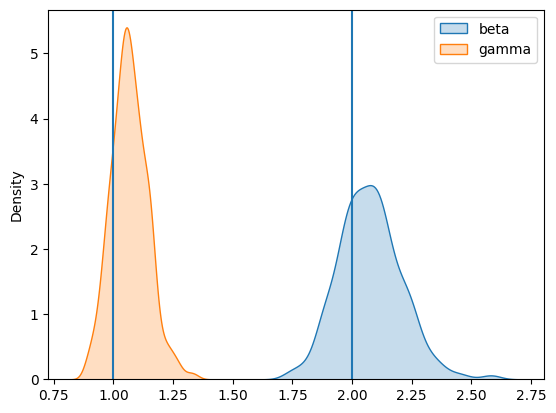}
\end{figure}

\begin{figure}[H]
\caption{Sampled 95\% credible intervals from PMCMC on SIR under-reported data with $p_{obs}$=0.05.}
\centering
\includegraphics[width=0.7\textwidth]{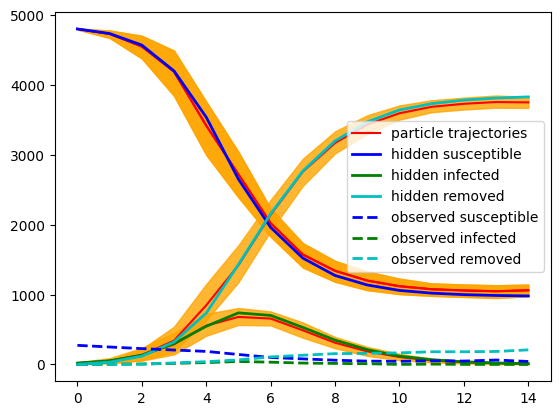}
\end{figure}

For the 0.025 $p_{obs}$, the posterior mean of $\beta$ was 2.025 with a high posterior density interval of (1.786, 2.288), while the posterior mean of $\gamma$ was 1.021 with a high posterior density interval of (0.89, 1.167). The posterior mean squared error of $\beta$ and $\gamma$ were 0.017166 and 0.005473 respectively.

\begin{figure}[H]
\caption{Density plot of posterior samples of $\beta$ and $\gamma$ from PMCMC on SIR under-reported data with $p_{obs}$=0.025.}
\centering
\includegraphics[width=0.7\textwidth]{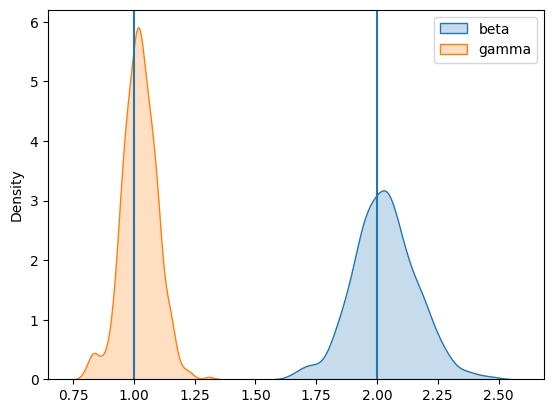}
\end{figure}

\begin{figure}[H]
\caption{Sampled 95\% credible intervals from PMCMC on SIR under-reported data with $p_{obs}$=0.025.}
\centering
\includegraphics[width=0.7\textwidth]{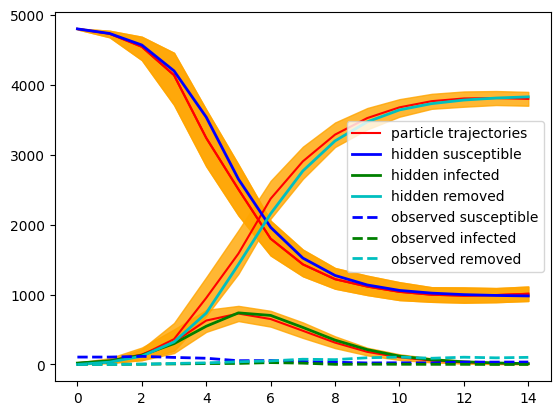}
\end{figure}

For the 0.01 $p_{obs}$, the posterior mean of $\beta$ was 2.035 with a high posterior density interval of (1.719, 2.418), while the posterior mean of $\gamma$ was 1.015 with a high posterior density interval of (0.787, 1.245). The posterior mean squared error of $\beta$ and $\gamma$ were 0.032917 and 0.014551 respectively.

\begin{figure}[H]
\caption{Density plot of posterior samples of $\beta$ and $\gamma$ from PMCMC on SIR under-reported data with $p_{obs}$=0.01.}
\centering
\includegraphics[width=0.7\textwidth]{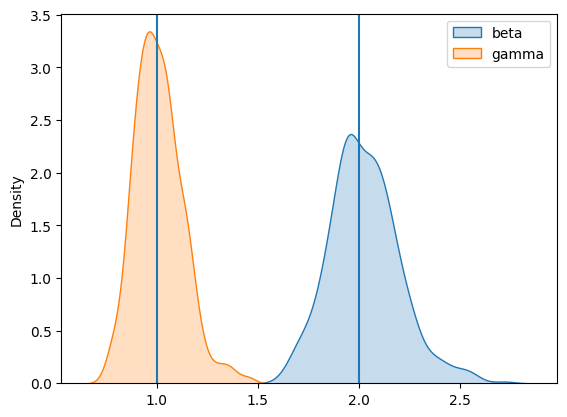}
\end{figure}

\begin{figure}[H]
\caption{Sampled 95\% credible intervals from PMCMC on SIR under-reported data with $p_{obs}$=0.01.}
\centering
\includegraphics[width=0.7\textwidth]{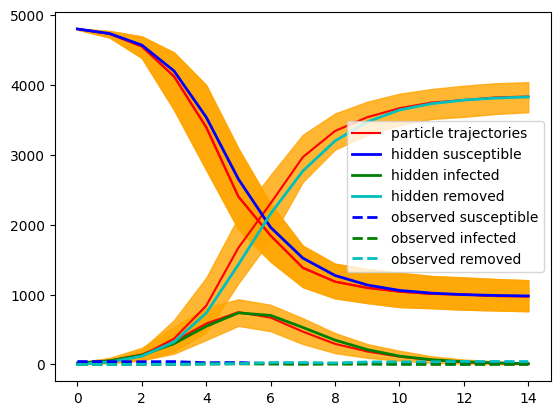}
\end{figure}

From the results above we can deduce that while decreasing the probability of observation in the under-reported dataset, we still achieve good results. Similarly to the tests run in Subsection \ref{sec:noise_level} on increasing the noise, the posterior variance of the results seems to increase, while the posterior means remain very close to their true values. The posterior mean squared errors seem to generally be increasing when we reduce $p_{obs}$ and the variability of the sampled parameters and trajectories increases, as it can be seen from their credible intervals. This can be attributed to the reduction of information and data-points read by the model every time we reduce the probability of observation. Yet, fairly good results are still achieved, even with this limited amount of data.

\subsubsection{Time-points observed level}

The effect of reducing the time-points observed in our predictions is additionally investigated in this subsection. In epidemiology it is essential to gather information about the disease as early as possible, so using a simple SIR model with noisy data I try to find at which point during the disease we would be able to make accurate predictions about the disease's dynamics. I use the same dataset as described in Subsection \ref{sec:noisy_sir}, which had $n_{ratio}$ of 0.01, and I test reducing the time-points from 15 originally to 11, 7 and 3. I run one chain for each case with 6,000 steps each with a burn-in of 1,000 samples and a thinning of 10 samples for the visualisations only.

When using the 11 first time-points, the posterior mean of $\beta$ was 1.948 with a high posterior density interval of (1.441, 2.504), while the posterior mean of $\gamma$ was 0.958 with a high posterior density interval of (0.521, 1.424). The posterior mean squared error of $\beta$ and $\gamma$ were 0.082036 and 0.063657 respectively.

\begin{figure}[H]
\caption{Density plot of posterior samples of $\beta$ and $\gamma$ from PMCMC on SIR noisy data with $n_{ratio}$=0.01 and 11 observed time-points.}
\centering
\includegraphics[width=0.7\textwidth]{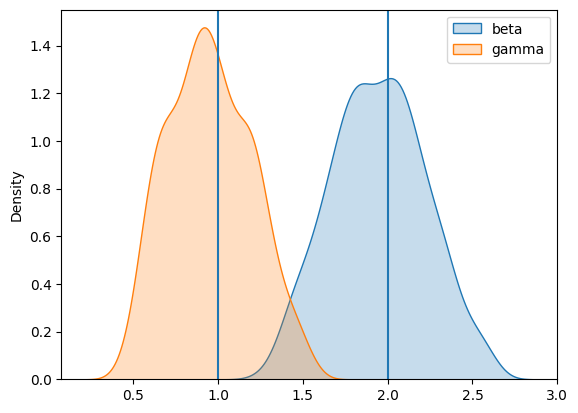}
\end{figure}

\begin{figure}[H]
\caption{Sampled 95\% credible intervals from PMCMC on SIR noisy data with $n_{ratio}$=0.01 and 11 observed time-points.}
\centering
\includegraphics[width=0.7\textwidth]{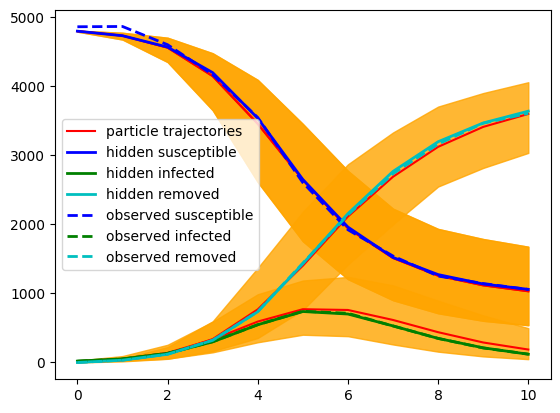}
\end{figure}

When using the 7 first time-points, the posterior mean of $\beta$ was 2.308 with a high posterior density interval of (1.475, 3.326), while the posterior mean of $\gamma$ was 1.18, with a high posterior density interval of (0.354, 1.891). The posterior mean squared error of $\beta$ and $\gamma$ were 0.357116 and 0.218229 respectively.

\begin{figure}[H]
\caption{Density plot of posterior samples of $\beta$ and $\gamma$ from PMCMC on SIR noisy data with $n_{ratio}$=0.01 and 7 observed time-points.}
\centering
\includegraphics[width=0.7\textwidth]{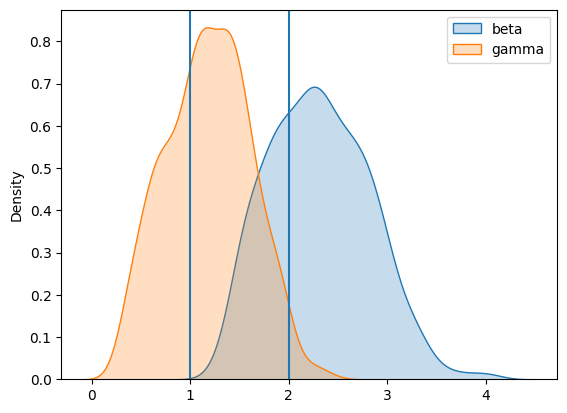}
\end{figure}

\begin{figure}[H]
\caption{Sampled 95\% credible intervals from PMCMC on SIR noisy data with $n_{ratio}$=0.01 and 7 observed time-points.}
\centering
\includegraphics[width=0.7\textwidth]{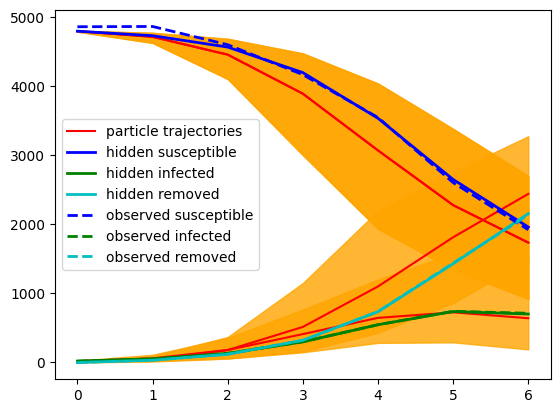}
\end{figure}

When using the 3 first time-points, the posterior mean of $\beta$ was 16.039 with a high posterior density interval of (1.072, 36.931), while the posterior mean of $\gamma$ was 12.944 with a high posterior density interval of (0.102, 33.774)). The posterior mean squared error of $\beta$ and $\gamma$ were 316.852791 and 244.141619 respectively.

\begin{figure}[H]
\caption{Density plot of posterior samples of $\beta$ and $\gamma$ from PMCMC on SIR noisy data with $n_{ratio}$=0.01 and 3 observed time-points.}
\centering
\includegraphics[width=0.7\textwidth]{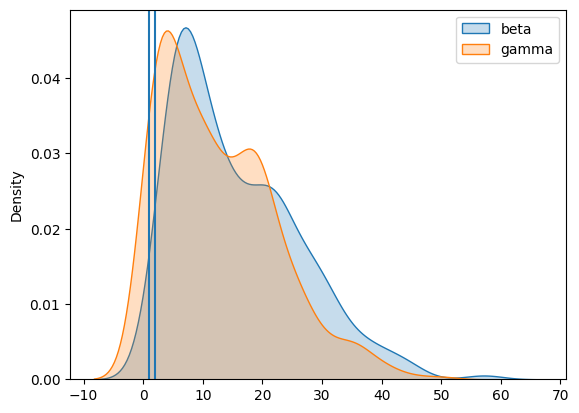}
\end{figure}

\begin{figure}[H]
\caption{Sampled 95\% credible intervals from PMCMC on SIR noisy data with $n_{ratio}$=0.01 and 3 observed time-points.}
\centering
\includegraphics[width=0.7\textwidth]{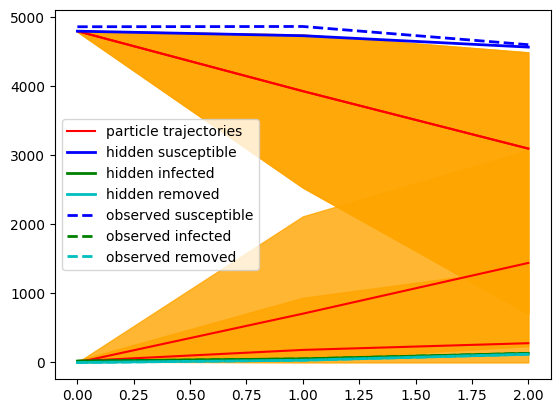}
\end{figure}

We deduce that when using the 11 first time-points we still get decent results, although the posterior variance increases significantly and the credible intervals of the sampled trajectories get much wider, and that could be since the peak of the disease has passed by this point and it is still possible to infer the parameters' posterior distributions from the data available. When reducing the time-points to 7, we notice that there is another increase in the posterior covariance of the parameters and the parameters are being overestimated. This could be because of the very limited amount of time-points. When only using the first 3 time-points, the posterior estimates of the parameters are overestimated and are very far from the true values, while their posterior variances are significantly increased, to a point that it seems as if the model cannot possibly converge to the truth. This is probably due to the lack of identifiability in the dataset -for this amount of information there are possibly many different parameters that could explain the observed data. That could be due the limited time-points and specifically for not having the peak of the disease included in the data provided, meaning that in general the data could result into very different epidemics.

\section{Discussion}

In this project the use of Particle Markov Chain Monte Carlo in stochastic epidemic modelling is investigated, using multiple simulated datasets and hidden Markov Chains for different possible scenarios. Additionally, an adaptive approach to modelling the posterior covariance of the parameters is explored in order to model scenarios where the information in the data read by the model would sequentially reduce. The algorithm was able to retrieve the estimated posterior densities, even with a very limited amount of information on the data. For models with few parameters very good posterior densities were received, centered at the true values, and credible intervals around the trajectories that were very accurate. When reducing the amount of data observed through the hidden Markov Chains, the uncertainty of the results was evidently increasing. Yet, very accurate posterior estimates were still received, when the identifiability requirement was not violated. These results are promising, since in a real world setting the uncertainty of the observed cases of a disease is expected to be very close to the ones modelled here. Thus, any parties interested in getting some of the characteristics underlying an epidemic's dynamics could use this algorithm to reach some estimates quickly, with a minimum amount of data that are accessible to anyone.

While the algorithm can be very useful in certain settings, it is also not without limitations. One such limitation is that the algorithm is very computationally intensive. MCMC by nature is a class of algorithms that due to Monte Carlo, requires generating a large amount of samples and when extending it to include the particle filter, which generates a large amount of particles at each step, it requires significant computational resources. In this project only small datasets were simulated and simple models used, but it was still difficult to get an acceptance rate of more than 20\%, due to the complicated nature of the epidemic models. Thus, it requires an even larger sample than those used in simple MCMC algorithms to achieve convergence and achieve an uncorrelated estimate of the posterior density. In addition, increasing the dimensions of the parameters makes the algorithm even more computationally intensive. During this project it was also attempted to fit a model with multiple subgroups, which had different rates of contact and infection between the subgroups. It proved to be challenging to deliver good estimates of the posterior covariance and achieve convergence, so due to time constraints the experiment was dropped. This could be due to the use of the Metropolis-Hastings algorithm and perhaps a different MCMC algorithm would be better suited for a large number of parameters. In recent years new MCMC algorithms have been proposed that are better suited for multi-dimensional inference\cite{durmus-2017} and for ODE based models\cite{valderrama-bahamondez-2019} which could be an interesting extension to the work presented here by implementing the particle filter in combination with those algorithms.

In relation to possible extensions to this work, a couple of topics were aimed to be included in this work, but this has not been feasible due to the limited time-frame. It would be interesting to try to fit an SIR like model with the model trained only in a single compartment. This should probably be the infectious compartment, which could be very useful, since in real world settings the infectious count data  are sometimes the only available. In addition, it would be intriguing to try to fit a model with multiple subgroups, but with observing only the totals by compartment, to reduce further the amount of information needed. Possible extensions that could be of interest, would be to try to fit more complicated models, like the SIS\cite{nakamura-2019} or SEIRS\cite{bjrnstad-2020}, or models with changing populations due to demographics\cite{martcheva-2015}. These scenarios would add a layer of complexion due to the infectious curve having multiple modes and being harder to model, which is necessary when dealing with diseases that have an element of seasonality, like the influenza.

\printbibliography
\end{document}